\documentclass[11pt,a4paper]{article}
\pdfoutput=1
\usepackage{epsfig}
\usepackage{subcaption}
\usepackage{graphicx}
\usepackage[english]{babel}
\usepackage[utf8]{inputenc}
\usepackage{mathtools}
\usepackage{amsmath}    
\usepackage{jheppub}
\usepackage[english]{babel}
\usepackage[utf8]{inputenc}
\usepackage{amsmath}
\usepackage{graphicx}
\usepackage{mathtools}
\usepackage{braket}

\title{Vacua and correlators in hyperbolic de Sitter space}
                                           
\author[a]{Fotios V. Dimitrakopoulos}
\author[a]{,Laurens Kabir}
\author[a]{,Benjamin Mosk}
\author[b]{,Maulik Parikh}
\author[a]{and Jan Pieter van der Schaar}
\affiliation[a]{Delta Institute for Theoretical Physics\\
IOP and GRAPPA, Universiteit van Amsterdam,\\
Science Park 904, 1090 GL Amsterdam, Netherlands}
\affiliation[b]{Department of Physics and Beyond: Center for Fundamental Concepts in Science\\ Arizona State University, Tempe, Arizona 85287, USA\\}
\emailAdd{F.Dimitrakopoulos@uva.nl}
\emailAdd{L.R.Kabir@uva.nl}
\emailAdd{B.Mosk@uva.nl}
\emailAdd{Maulik.Parikh@asu.edu}
\emailAdd{J.P.vanderSchaar@uva.nl}
\abstract{We study the power-- and bi--spectrum of vacuum fluctuations in a hyperbolic section of de Sitter space, comparing two states of physical interest: the Bunch-Davies and hyperbolic vacuum. We introduce a one--parameter family of de Sitter hyperbolic sections and their natural vacua, and identify a limit in which it reduces to the planar section and the corresponding Bunch--Davies vacuum state. Selecting the Bunch--Davies vacuum for a massless scalar field implies a mixed reduced density matrix in a hyperbolic section of de Sitter space. We stress that in the Bunch--Davies state the hyperbolic de Sitter $n$-point correlation functions have to match the planar de Sitter $n$-point correlation functions. The expressions for the planar and hyperbolic Bunch--Davies correlation functions only appear different because of the transformation from planar to hyperbolic coordinates. Initial state induced deviations from the standard inflationary predictions are instead obtained by considering the pure hyperbolic vacuum, as we verify explicitly by computing the power-- and bi--spectrum. For the bi--spectrum in the hyperbolic vacuum we find that the corrections as compared to the standard Bunch--Davies result are not enhanced in specific momentum configurations and strongly suppressed for momenta large compared to the hyperbolic curvature scale. We close with some final remarks, in particular regarding the implications of these results for more realistic inflationary bubble scenarios.}
\begin{document}
\maketitle
\flushbottom


\section{Introduction}

Cosmological observations point to a primordial universe that can be effectively described by an approximate de Sitter phase. The details of this inflationary phase are to a large extent still unknown, but the most recent Planck data does rule out a large fraction of parameter space, giving us some hints about the underlying physics \cite{Planck:2013}. From a theoretical point of view it is probably fair to say that inflation is poorly understood. Indeed, few convincing theoretical constraints on the inflationary parameter space exist that would identify natural, UV consistent, models. Trying to embed inflation into a fundamental description like string theory is notoriously difficult, but the last decade has seen considerable phenomenological progress in that direction (for a thorough discussion see \cite{BauMcA:2014} and references therein). The absence of a guiding principle which is able to rule out a significant fraction of inflationary models, combined with the attractive features of eternal inflation, has fueled the idea that perhaps our inflationary universe is just one realization in a huge landscape of bubble universes that are continuously being produced as a consequence of a stochastically varying scalar field during a phase of eternal inflation\cite{Vilenkin:1983, Susskind:2003}. 

This exotic possibility makes the general prediction that the spatial sections in our universe should be hyperbolic on the largest scales that start probing the boundary of the bubble. Clearly this general prediction is hard, if not impossible, to verify because the primordial inflationary expansion typically redshifts the negative curvature scale far beyond the observable universe \cite{Guth:2012}, although in the inflationary landscape relatively short phases of slow-roll inflation might be preferred \cite{Freivogel:2006}, which could lead to observable consequences in the CMB temperature correlations at low multipoles \cite{YamLinNarSasTan:2011}. More particular predictions include a bubble universe that might, under fortuitous conditions, provide an explanation for the low power anomaly at low $l$ due to the steepening of the slope right after penetration of the barrier \cite{BouHarSen:2013}. Another potential consequence of observational interest is that bubble universes can in principle collide \cite{Kleban:2009, Kleban:2011, Kleban:2012}, which could leave definite non-isotropic signatures in the Cosmic Microwave Background sky. Unfortunately the chance for a collision to have taken place in our past is small and model-dependent, and therefore not seeing this effect will not be able to rule out a multiverse origin \cite{Peiris:2011}. 

Recently, another rather generic consequence of a multiverse origin has been explored \cite{Agullo:2010, Komatsu:2013, Kanno:2014}. In the context of an inflationary landscape one would expect the initial vacuum state for quantum fluctuations in a single inflationary bubble to be entangled with the rest of the eternally inflating universe, leading to a mixed state inside the bubble. Since the Cosmic Microwave Background temperature anisotropies (as well as the large scale structure distribution) is probing the statistics of these inflationary quantum fluctuations, one could imagine uncovering evidence in favor of a mixed initial state that would support the idea that our universe originated from false vacuum decay. This idea warrants a careful study of the actual (observational) potential to constrain departures away from a standard pure initial state for inflationary quantum fluctuations and how these departures relate to the global vacuum state of eternal inflation. 

In this work we present a first step, triggered by some recent work in this direction \cite{Kanno:2014}, in clarifying the connection between the vacuum state of the false, eternally inflating, vacuum and potential departures from the standard Bunch--Davies vacuum state in a hyperbolic bubble. As a starting point we identify a limit that connects hyperbolic coordinates and its naturally associated vacuum to the de Sitter invariant Bunch--Davies state on planar sections. Following up on older work in the context of open inflation, after selecting the global Bunch--Davies vacuum we then use the mixed reduced density matrix defined in a single hyperbolic coordinate patch of de Sitter space to explicitly show that the statistics of inflationary quantum fluctuations are indistinguishable from the standard planar Bunch--Davies predictions at late times and we explain why this should not come as a surprise. The mixed nature of the initial state in hyperbolic de Sitter space in fact ensures that all predictions match with those of the pure Bunch--Davies state in the planar de Sitter space that one started out with. The mixed nature of the initial hyperbolic vacuum state therefore, in this case, does not imply observable departures from the standard planar Bunch-Davies predictions for the power- and bi-spectrum. Selecting instead the pure hyperbolic vacuum does lead to differences that are however strongly suppressed in the curvature scale, as we again show explicitly by computing the power-- and bi--spectrum for scalar field vacuum fluctuations. We end with a discussion on the implications in the context of (open) inflationary models and some remaining open questions that we hope to return to in future work.


\section{A family of de Sitter hyperbolic sections and their vacua}

We will start by constructing a limit in the family of de Sitter hyperbolic sections that reduces to the planar description of de Sitter space. This allows us to explicitly see the distinction between the natural vacuum state on a generic hyperbolic section and the Bunch-Davies vacuum state as defined using planar coordinates, which can be understood as a singular (but well-defined) limit of the natural hyperbolic vacuum.   

\subsection{A generalized hyperbolic embedding}

\begin{figure}[ht]
\centering
\begin{minipage}[t]{\textwidth}
\centering
\includegraphics[width=0.5\textwidth]{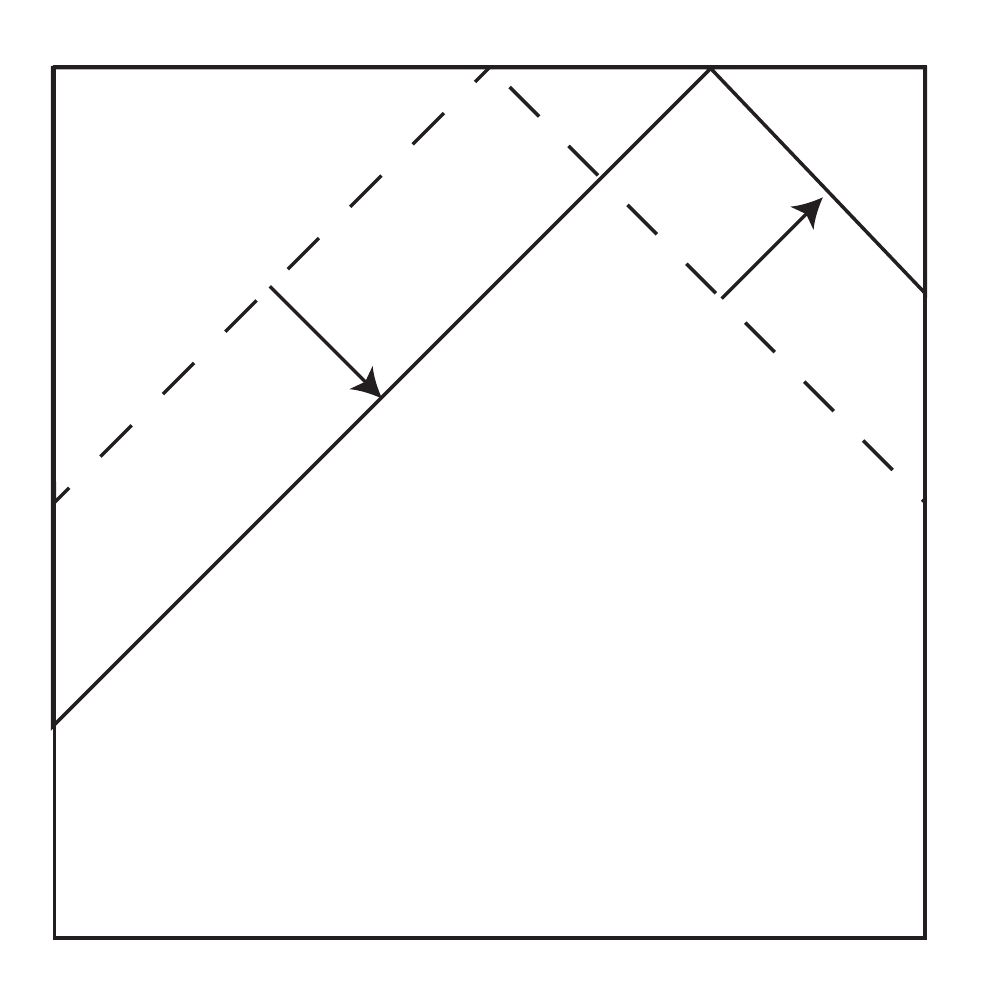}\caption{Conformal diagram of $dS_4$ with the left- and right-hyperbolic patch as the upper-left resp. upper-right triangles. The dashed line is the unboosted situation $\gamma=0$. For finite $\gamma$ (solid line), we see that that nucleation time of the left bubble gets pushed to earlier times, and vice versa for the right bubble. In the limit of $\gamma\rightarrow \infty$, we can see that the left bubble will cover the entire upper-left triangle of the conformal diagram, coinciding with the planar patch.}\label{fig:hyperbolicboost}
\end{minipage}
\end{figure}

Let us remind the reader that $4$-dimensional de Sitter space can be defined as the embedding surface in $5$-dimensional flat space defined by
\begin{equation}
-X_0^2 + X_1^2 + X_2^2 + X_3^2 + X_4^2 = 1 \, 
\label{def-ds}
\end{equation}
where the de Sitter curvature length scale has been normalized to one. This embedding equation is clearly invariant under $SO(1,4)$ transformations, corresponding to the isometry group of $dS_4$. 
For our purposes we will be interested in two different coordinate sets on this embedding surface that both belong to the class of isotropic and homogeneous FRLW spaces. The standard coordinate set used for describing inflation is the planar one, identifying flat spatial sections. In what follows, we will suppress the two (spatial) coordinates $X_3$ and $X_4$ for purposes of efficiency, effectively suppressing an $S^2$ in the de Sitter space. The planar coordinates are defined as
\begin{eqnarray}
X_0 + X_1 &=& e^{t_p} \,  \nonumber \\
X_0 - X_1 &=& \left( r_p^2 \, e^{2t_p} -1 \right) e^{-t_p} \, \\
X_2 &=& r_p \, e^{t_p} \,  \nonumber
\label{planar}
\end{eqnarray}
leading to the well-known planar expression for the induced de Sitter metric
\begin{equation}
ds^2 = -dt_p^2 + e^{2t_p} \left[ dr_p^2 + r_p^2 \, d\Omega_2^2 \right] \, 
\label{planarmetric}
\end{equation}
with $-\infty < t_p < +\infty$ and $r>0$, and we reinserted the $S^2$ part. Since $X_0+X_1 \geq 0$ these coordinates only cover the upper half diagonal part in the $X_0$ versus $X_1$ plane. Besides the obvious $SO(3)$ isometries, the boost symmetries of the embedding space are realized on the planar metric as an isometry involving a particular combination of time translation and spatial scaling\footnote{More precisely, it corresponds with the isometry $t\rightarrow t+\gamma$ and $r\rightarrow e^{-\gamma}r$ of the planar de Sitter metric.}. Because inflation redshifts away any existing spatial curvature present initially, this coordinate set should be an excellent approximation to derive the late-time effects of a sustained phase of cosmological inflation. Nevertheless, one could imagine a situation where our universe has originated from a tunneling event out of an eternally inflating false vacuum\footnote{Moreover, in the nineties models of open inflation were of particular interest, independent of whether their origin was due to tunneling \cite{BucGolTur:1994, LytSte:1990}.}. The nucleated bubble would have negatively curved spatial sections \cite{ColLuc:1980, Gott:1982}, leading to the hyperbolic coordinate set
\begin{eqnarray}
X_0 + X_1 &=& \cosh{t_h} +\sinh{t_h} \, \cosh{r_h} \,  \nonumber \\
X_0 - X_1 &=& - \cosh{t_h} +\sinh{t_h} \, \cosh{r_h} \, \\
X_2 &=& \sinh{t_h} \, \sinh{r_h} \, . \nonumber
\label{hyperbolic}
\end{eqnarray}
With these identifications the induced hyperbolic de Sitter metric reads
\begin{equation}
ds^2 = -dt_h^2 + \sinh{t_h}^2 \left[ dr_h^2 + \sinh^2{r_h} \, d\Omega_2^2 \right] \,  
\label{hyperbolicmetric}
\end{equation}
where $0 \leq t_h<+\infty$ and $r_h>0$, and as before we included the full $S^2$ that was left out in the embedding coordinate identification. The coordinate singularity at $t_h=0$ can be interpreted in the context of false vacuum decay as the creation of the open inflationary bubble. Note that $t_h=0$ corresponds to $X_0=0$ (and $X_1=1$, $X_2=0$): the bubble nucleation time from the point of view of the embedding space. The spatial sections correspond to constant negative curvature slices that exhibit an $SO(1,3)$ isometry. We will in fact be interested in a one-parameter generalization of this hyperbolic coordinate embedding, obtained by boosting in the $X_0$--$X_1$ plane of the $5$--dimensional embedding space. Combined with rotations these transformation allow one to move the `nucleation' time of the hyperbolic bubble to any specific point on the embedding surface. Just performing a Lorentz boost in the $X_0$--$X_1$ plane will change the nucleation time (and position in $X_1$), which yields the following generalized hyperbolic coordinate set
\begin{eqnarray}
X_0 + X_1 &=& e^{-\gamma} \, \left[ \cosh{t_h} +\sinh{t_h} \, \cosh{r_h} \right] \,  \nonumber \\
X_0 - X_1 &=& e^{\gamma} \, \left[ -\cosh{t_h} +\sinh{t_h} \, \cosh{r_h} \right]\, \\
X_2 &=& \sinh{t_h} \, \sinh{r_h} \,  \nonumber
\label{gammahyperbolic}
\end{eqnarray}
where $\gamma$ is the boost parameter. This generalized hyperbolic solution of the embedding equation will of course lead to the same induced metric, but the nucleation time and position of the associated bubble in the embedding space have now shifted to $X_0 = -\sinh{\gamma}$ and $X_1= \cosh{\gamma}$ respectively. Moreover, since $t_h \geq 0$ one finds that $X_0 + X_1 \geq e^{-\gamma}$, restricting the hyperbolic section to the upper right diagonal part in the $X_0$ versus $X_1$ plane, which overlaps with, but for any finite $\gamma$ is smaller than, the part of de Sitter covered by planar coordinates. This is depicted in figure \ref{fig:hyperbolicboost}. One can verify that in the limit of infinite $\gamma$ the planar and hyperbolic coordinates cover the same region of de Sitter space, which is consistent with the observation that in this limit the hyperbolic nucleation time in the embedding space is shifted to $X_0 \rightarrow -\infty$. 

The generalization of the hyperbolic coordinate set introduced above allows us to explicitly relate the planar and hyperbolic sections of de Sitter space. Since the two coordinate sets cover the same region in the $\gamma \rightarrow \infty$ limit, there should exist a one-to-one mapping between the coordinates in that limit. More precisely, we would like to introduce a new set of hyperbolic coordinates that are to be kept fixed in the limit $\gamma \rightarrow \infty$, and that in the limit exactly reproduce the planar coordinate embedding solution. Note that any (constant) shift or rescaling of the hyperbolic embedding coordinates is still a solution of the embedding equation, but will change the expression for the induced metric. Since we expect the range of the hyperbolic time coordinate to be extended to $-\infty$ and the negative curvature to be scaled away, we redefine 
\begin{equation}
\tilde t_h \equiv t_h -\gamma ~ ; ~ \tilde r_h \equiv \frac{1}{2} r_h \, e^\gamma \, . 
\label{fixedcoord}
\end{equation}
This leaves us with the following generalized hyperbolic solution to the embedding equation
\begin{eqnarray}
X_0 + X_1 &=& e^{-\gamma} \, \left[ \cosh{(\tilde t_h +\gamma)} +\sinh{(\tilde t_h + \gamma)} \, \cosh{(2\tilde r_h \, e^{-\gamma})} \right] \,  \nonumber \\
X_0 - X_1 &=& - e^{\gamma} \, \left[ \cosh{(\tilde t_h + \gamma)} - \sinh{(\tilde t_h +\gamma)} \, \cosh{(2\tilde r_h \, e^{-\gamma})} \right]\, \\
X_2 &=& \sinh{(\tilde t_h +\gamma)} \, \sinh{(2\tilde r_h \, e^{-\gamma})} \,  \nonumber
\label{gammahyperbolictilde}
\end{eqnarray}
where $-\gamma \leq \tilde t_h < +\infty$. For finite $\gamma$ the shift in hyperbolic time and the rescaling of the hyperbolic radius (or equivalently the inverse rescaling of hyperbolic momentum) does obviously not affect any hyperbolic patch observables, but it does allow one to analyze the infinite boost limit in a simple and useful way. The induced hyperbolic metric now reads 
\begin{eqnarray}
ds^2 &=& -d\tilde t_h^2 + \sinh{(\tilde t_h + \gamma)}^2 \left[ 4 e^{-2\gamma} \, d\tilde r_h^2 + \sinh{(2 \tilde r_h e^{-\gamma})}^2 \, d\Omega_2^2 \right] \label{hyperbolictildemetric} \\
&\overset{\gamma \rightarrow \infty }= & -d\tilde t_h^2 + e^{2 \tilde t_h} \left[d\tilde r_h^2 + \tilde r_h^2 \, d\Omega_2^2 \right] \, . \nonumber
\end{eqnarray}
In the second line we performed the limit $\gamma \rightarrow \infty$, keeping $\tilde t_h$ and $\tilde r_h$ fixed, showing that (\ref{gammahyperbolictilde}) exactly reduces to the planar embedding solution (\ref{planar}). Note that all the $\gamma$ dependence in the induced metric is removed and one is left with precisely the planar line-element (\ref{planarmetric}) in terms of the coordinates $\tilde t_h$ and $\tilde r_h$. 

For any finite boost parameter $\gamma$ global de Sitter space is covered by two (adjacent) hyperbolic sections, see figure \ref{fig:hyperbolicboost}. The other hyperbolic embedding can be obtained by changing the sign of $X_1$, resulting in the interchange of the expressions for $X_0+X_1$ and $X_0-X_1$ in (\ref{hyperbolic}).
Acting with the same boost on this second hyperbolic embedding results in the opposite effect, moving the nucleation time to $X_0 \rightarrow +\infty$. The opposite minus infinity boost should instead reduce to another planar section (with $X_0+X_1$ and $X_0-X_1$ in (\ref{planar}) interchanged), suggesting that the redefined coordinates in this case should read 
\begin{equation}
\tilde t_h \equiv t_h +\gamma ~ ; ~ \tilde r_h \equiv \frac{1}{2} r_h \, e^{-\gamma} \, . 
\label{fixedcoord2}
\end{equation}
Putting this together we obtain for the adjacent hyperbolic section the following generalized embedding
\begin{eqnarray}
X_0 + X_1 &=& - e^{-\gamma} \, \left[ \cosh{(\tilde t_h - \gamma)} - \sinh{(\tilde t_h -\gamma)} \, \cosh{(2\tilde r_h \, e^{\gamma})} \right]
\,  \nonumber \\
X_0 - X_1 &=& e^{\gamma} \, \left[ \cosh{(\tilde t_h - \gamma)} + \sinh{(\tilde t_h -\gamma)} \, \cosh{(2\tilde r_h \, e^{\gamma})} \right]
\, \\
X_2 &=& \sinh{(\tilde t_h -\gamma)} \, \sinh{(2\tilde r_h \, e^{\gamma})} \,  \nonumber
\label{gammahyperbolictilde2}
\end{eqnarray}
where now $\gamma \leq \tilde t_h < +\infty$. The induced hyperbolic metric in this case is obtained by just replacing $\gamma$ with $-\gamma$ in (\ref{hyperbolictildemetric}). By construction the limit $\gamma \rightarrow \infty$ should instead collapse and in a sense remove the adjacent hyperbolic section. Clearly, in the opposite $\gamma \rightarrow -\infty$ the roles of the two hyperbolic sections are reversed.   

\begin{figure}[h!]
\centering
\begin{minipage}[t]{0.48\textwidth}
\centering
\includegraphics[width=\textwidth]{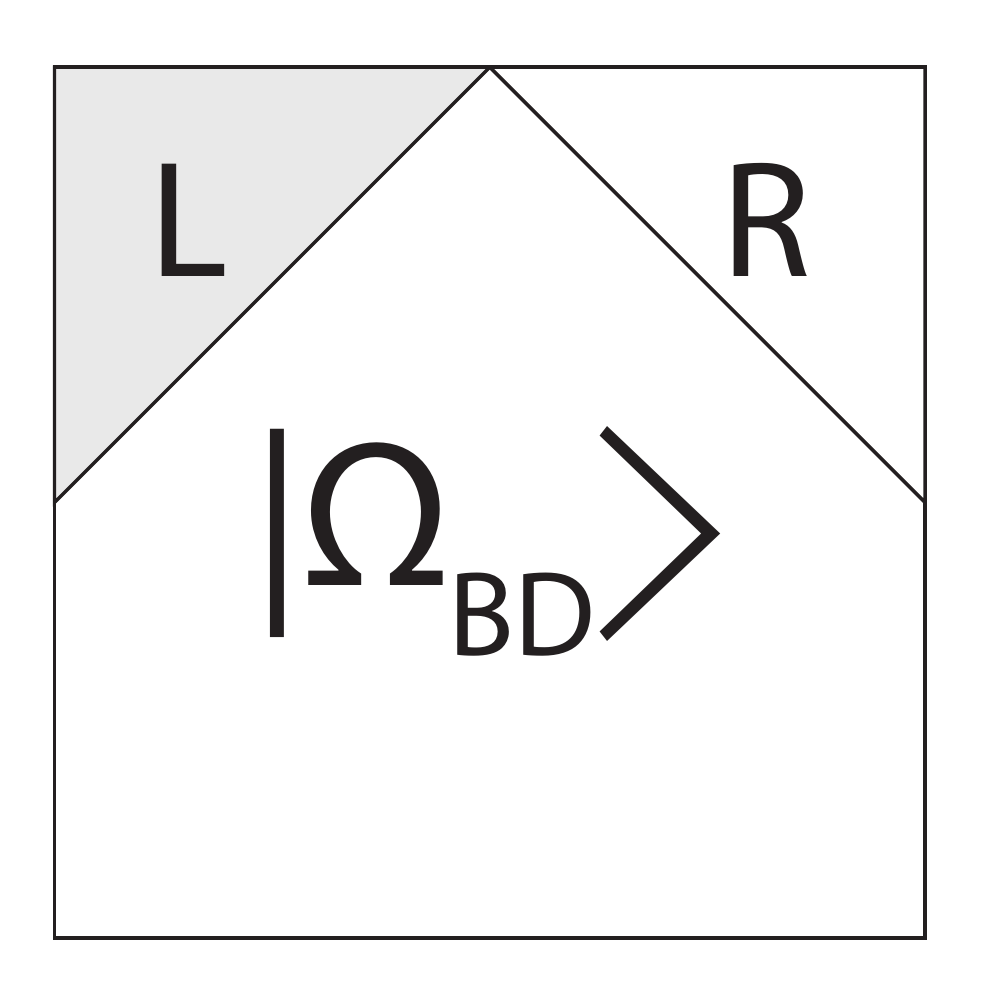}
\caption{Conformal diagram of $dS_4$ with the left- and right-hyperbolic patch shown. As neither patch contains a Cauchy slice of the full $dS_4$, restricting the Bunch-Davies vacuum to one of them will yield a mixed state.}\label{fig:hyperbolicpatches}
\end{minipage}
\quad
\begin{minipage}[t]{0.48\textwidth}
\centering
\includegraphics[width=\textwidth]{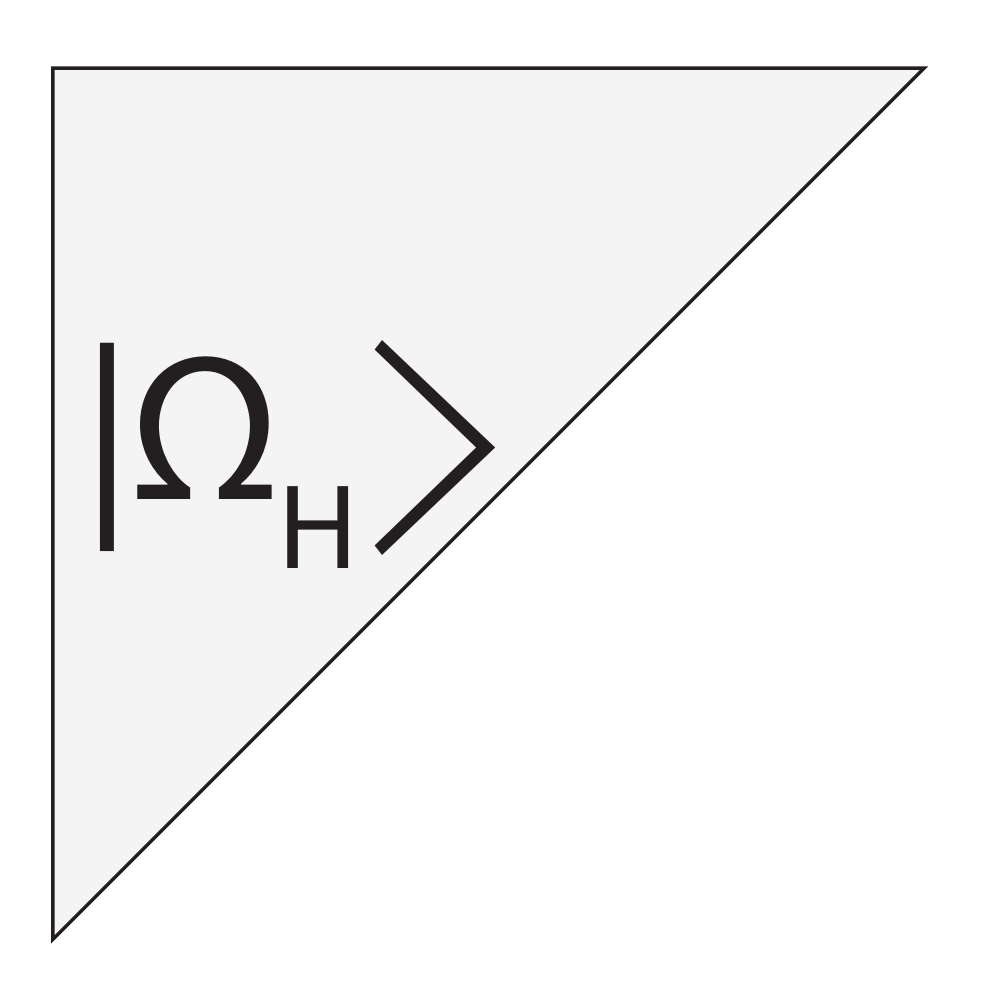}
\caption{An observer confined to live in the left-hyperbolic patch (a bubble universe) can define his own pure hyperbolic vacuum. This state will differ significantly from the mixed state resulting from a restriction of the Bunch-Davies state to this bubble.}\label{fig:lefthyperbolicpatch}
\end{minipage}
\end{figure}

Having established this explicit relation between hyperbolic and planar coordinates, we can now use it to better understand and connect their respective vacua, which should be different for any finite value of $\gamma$. In particular, the planar Bunch--Davies state is known to be equivalent to the unique and de Sitter invariant Euclidean vacuum\footnote{The invariance of the Bunch--Davies vacuum under de Sitter isometries strictly speaking fails for massless fields, but since this subtlety does not affect our results we will ignore it from now on.}. On the other hand, any pure hyperbolic vacuum state is defined on a negatively curved spatial slice that is not a de Sitter Cauchy surface. This means that the Bunch--Davies state in a single hyperbolic patch can only be described by an appropriately defined mixed state; see figures \ref{fig:hyperbolicpatches} and \ref{fig:lefthyperbolicpatch}. The mixed state defined on one of the two (conjugate) hyperbolic sections reproducing the Bunch--Davies state was first constructed in \cite{Sasaki:1995} and was subsequently used in \cite{MalPim:2012} to compute the reduced density matrix and the corresponding entanglement entropy for a single hyperbolic section. 

One application of the one-parameter family of hyperbolic de Sitter foliations is that one can confirm that the natural choice for a hyperbolic vacuum reduces to the planar Bunch--Davies state in the limit $\gamma \rightarrow \infty$. Secondly, one could attempt to generalize the entangled expression for wavefunctions of the Bunch--Davies state, with support on both the left and right hyperbolic section, and work out its dependence on the embedding boost parameter. In the $\gamma \rightarrow \infty$ limit this should reduce to the pure planar Bunch--Davies state, implying that the reduced density matrix carries some non-trivial $\gamma$ dependence to make sure the associated entanglement entropy vanishes in the strict $\gamma \rightarrow \infty$ limit.

To summarize, we established that the infinite boost limit of a hyperbolic de Sitter patch (and as a consequence also its corresponding vacuum state) reduces to the planar de Sitter patch (and the Bunch--Davies vacuum). This appears to be similar to an observation made in \cite{GreParScha:2006} where the static vacuum, understood as the empty state for a corresponding free-falling observer, was also argued to reduce to the Bunch--Davies state in the infinite boost limit. Note that to each hyperbolic patch one can associate a free-falling observer in one of the two center regions in between the hyperbolic patches that never intersects either one of them. These time-like curves are indeed connected to each other by the same embedding space boosts \cite{ParVer:2004}. To complete the argument one needs to confirm that the static vacuum state associated to this free-falling observer is connected to the hyperbolic vacuum state. Note that (for $\gamma=0$) the center region in between the hyperbolic patches is usually covered by coordinates that are obtained from the hyperbolic coordinates as follows $t_h = i (t_C - \frac{\pi}{2})$ and $r_h = r_C + i \frac{\pi}{2}$, resulting in the following center region metric
\begin{equation}
ds^2= dt_C^2 + \cos{t_C}^2 \left[-dr_C^2 + \cosh{r_C}^2 d\Omega^2\right] \, ,
\end{equation}
where $r_C$ is now a time-like coordinate. Each of the two center regions clearly identifies a causal diamond belonging to the free-falling observer of interest. To make this explicit, one notices that the coordinate transformation $r_s \equiv \sin{t_C}$ and $t_S \equiv r_C$ indeed reproduces the static patch metric, upon ignoring the two-dimensional sphere\footnote{That the $S^2$ part does not reproduce the standard static patch expression can be understood by realizing that this static patch region is rotated by an angle $\pi/2$ with respect to the standard embedding. This affects the $S^2$ angles, which have to be transformed as well in order to obtain the complete static patch metric.}. This establishes a map from the hyperbolic to the static patch of a specific free-falling observer, relating their respective vacua and their behavior in the infinite (embedding) boost limit.

\subsection{The hyperbolic vacuum}\label{sectionhyperbolic}

After having established a limit to obtain the planar embedding and coordinates, let us now remind the reader of the standard positive frequency modes on a single hyperbolic patch \cite{Sasaki:1995}, as if it were the entire universe (see figure \ref{fig:lefthyperbolicpatch}).  The scalar wave equation for the hyperbolic patch of de Sitter \eqref{hyperbolicmetric} reads
\begin{equation}
\left[\frac{1}{\sinh^3 t}\partial_t \sinh^3 t \partial_t - \frac{1}{\sinh^2 t} \nabla_{\mathcal{H}^3}^2 + m^2 \right]\phi=0
\label{eq:hyperboliceom}
\end{equation}
where we defined the Laplacian on the three-hyperboloid
\begin{equation}
\nabla_{\mathcal{H}^3}^2=\frac{1}{\sinh^2 r}\partial_r (\sinh^2 r \partial_r)+\frac{1}{\sinh^2 r}\nabla_{\mathcal{S}^2}^2 .
\label{eq:hyperboliclaplacian} 
\end{equation}
A natural set of solutions to the hyperbolic equations of motion \eqref{eq:hyperboliceom} is given by
\begin{equation}\label{Hmodes}
 \frac{1}{\sinh(t)}P^{ip}_{\nu-\frac{1}{2}}\left(\cosh (t)\right)Y_{plm}(r,\Omega)
\end{equation}
where it is customary to define $\nu = \sqrt{\frac{9}{4}-\frac{m^2}{H^2}}$. The quantum numbers $l,m$ label the usual $SO(3)$ irreps, and together with the continuous quantum number $p$ it completely specifies the hyperbolic momentum. Furthermore, $P^{ip}_{\nu-\frac{1}{2}}$ are the associated Legendre functions of the second kind and the $Y_{plm}$ are the orthonormal eigenfunctions of the hyperbolic Laplacian \eqref{eq:hyperboliclaplacian}
\begin{equation}
\nabla_{\mathcal{H}^3}^2 Y_{plm}(r,\Omega)=-(1+p^2)Y_{plm}(r,\Omega).
\end{equation}
For $\nu > \frac{1}{2}$, there is in fact a supplementary set of solutions with $p = i(\nu-\frac{1}{2})$ \cite{Sasaki:1995}. These so-called ``supercurvature modes" will not be  of interest for the purposes that are considered here, where our main focus will be on potential signatures in the large (``subcurvature'') momentum limit. We refer to \cite{Cohn:1998} for an interesting account on the role and interpretation of these supercurvature modes.

Switching to conformal time $\eta$ we can write the metric of the hyperbolic slice as
\begin{equation}
 \begin{aligned}
ds^2 &= \sinh^2(t(\eta))\left(-d\eta^2+dr^2+\sinh^2(r)d\Omega_2^2 \right),
 \end{aligned}
\end{equation}
where $\eta = \ln (\tanh(\frac{t}{2}))$, or equivalently $\cosh t = -\frac{1}{\tanh \eta}$ and $-\infty<\eta<0$. In terms of the conformal time $\eta$ we find that in the far past $\eta \rightarrow -\infty$ and in the limit of large momenta $p \gg 1$ one obtains
\begin{equation}
 \begin{aligned}
P^{ip}_{1}\left(-\frac{1}{\tanh \eta}\right) \propto e^{-ip\eta}\left(1-\frac{i}{p \tanh \eta}\right) \rightarrow e^{-i p\eta},
 \end{aligned}
\end{equation}
so we can identify these mode functions with the ``natural hyperbolic vacuum": they define a state that is empty in the far past for large momenta, approaching the standard vacuum description in flat space. As expected, in the limit $\gamma \rightarrow \infty$ that we introduced in the previous section (\ref{fixedcoord2}) the mode functions reduce to the standard Bunch--Davies mode functions in flat slicing, explicitly connecting the hyperbolic and planar patch vacua in this limit 
\begin{equation}
 \begin{aligned}
  \lim_{\gamma \rightarrow \infty} P^{ip}_{1}(\cosh (\tilde{t}+\gamma)) &\propto e^{-ip\tilde{\eta}}\left(1-\frac{i}{p\tilde{\eta}}\right).
 \end{aligned}
\end{equation} 
For all the details we refer the reader to the appendix \ref{app:modefunctionscalar}, but it should be clear that the tildes on the coordinates in the above equation relate to the redefined hyperbolic coordinates that are kept fixed in the infinite boost limit. 
The mode functions (\ref{Hmodes}) must of course be properly normalized, enforcing $[\hat{b}_{plm},\hat{b}_{plm}^{\dagger}] = \delta_{ll'} \delta_{mm'} \delta(p-p')$, implying the following Klein-Gordon inner product 
\begin{equation}
 \begin{aligned}
\langle \phi_{plm},\phi_{plm}\rangle_{\text{KG}} &= \delta_{ll'} \delta_{mm'} \delta(p-p')
 \end{aligned}
\end{equation}
giving (see appendix \ref{KGnorm})
\begin{equation}\label{Pnorm}
 \begin{aligned}
N_{P^p}^2 &\equiv \langle P^{ip},P^{ip} \rangle_{\text{KG}} \\ 
&= \frac{2\sinh(\pi p)}{\pi}.
 \end{aligned}
\end{equation}

With the help of (\ref{Pnorm}) we can now express the field operator in a single hyperbolic patch as (keeping in mind that we are ignoring supercurvature modes)
\begin{equation}\label{Hexpansion}
 \begin{aligned}
\phi(t,r,\Omega) &= \int_0^\infty dp \sum_{l=0}^\infty \sum_{m=-l}^l\frac{1}{N_{P^p}}\frac{1}{\sinh(t)}\left(\hat{b}_{plm}P^{ip}_{\nu-\frac{1}{2}}\left(\cosh (t)\right)Y_{plm}(r,\Omega)+\text{h.c.}\right)
 \end{aligned}
\end{equation}
defining the natural hyperbolic vacuum state $|\Omega_H\rangle$ as
\begin{equation}\label{Hvacuum}
 \begin{aligned}
\hat{b}_{plm}|\Omega_H\rangle &= 0 &&\forall p,l,m.
 \end{aligned}
\end{equation}


This hyperbolic vacuum state can be understood as a natural choice in an isolated (stand-alone) open inflationary universe, as it is empty in the far past and reduces to the planar Bunch--Davies vacuum state in the infinite boost limit. As we have elaborated upon in the previous section, because a similar statement can be made for the vacuum in a static patch \cite{GreParScha:2006}, and because the region in between the two hyperbolic patches contains a causal diamond region, this state can be analytically continued to the static (empty) vacuum for a free-falling observer that never intersects the two adjacent hyperbolic de Sitter patches.  

Clearly this state is very different from the unique de Sitter invariant Bunch--Davies vacuum for generic $\gamma$, so one would expect anomalous behavior similar to what happens in the de Sitter static vacuum or the flat Rindler vacuum. To that end let us analyze the behavior of the energy momentum tensor in a (generic) hyperbolic vacuum. Note that for a flat Rindler wedge in lightcone coordinates $(u,v)$, there is a horizon at $u=0$ and the Fulling-Rindler vacuum $|0_{FR}\rangle$ corresponds to the empty state in a single wedge. In that case it is well-known that the $T_{uu}$ component of the energy momentum tensor (with the usual UV-divergence removed by subtracting the UV-divergent expectation value of $T_{uu}$ in the Minkowski vacuum $|0_M\rangle$) diverges as one approaches the horizon: $\langle T_{uu}\rangle_{FR} - \langle T_{uu}\rangle_{M} = -\frac{1}{48\pi}\frac{1}{u^2}$ in $1+1$ dimensions for $u>0$ (for a nice derivation of this result see \cite{Parentani:1993}).

A similar analysis can be done for the energy momentum tensor in the hyperbolic de Sitter patch, where the global de Sitter invariant vacuum state is now the Bunch-Davies vacuum $|\Omega_{BD}\rangle$. The obvious difference with the Rindler wedge is the absence of a timelike Killing vector. In addition, the $t=0$ surface is a (light-) cone, so a better analogy is with Milne space, to which the de Sitter hyperbolic section reduces for small $t$. In any case, we will use the same regularization procedure, restricting to the minimally coupled massless case $\nu = \frac{3}{2}$. The most convenient method to calculate components of the energy momentum tensor makes use of the Wightman function $G^{+}(x,x',t,t')$ and specifically we will look at the following contribution
\begin{equation}
 \begin{aligned}
  \langle (\partial_{\alpha}\phi)^2(x,t) \rangle &= \lim_{x',t'\rightarrow x,t}\partial_{\alpha}\partial_{\alpha}G(x,x',t,t').
 \end{aligned}
\end{equation}
The Wightman function for the Bunch-Davies state is well known, but here we use the expression in terms of an integral over the hyperbolic momentum $p$ as given in \cite{Sasaki:1995}. This allows us to consistently regulate the UV-divergence of $\langle T_{\mu \nu}\rangle$ in the two states of interest. In appendix \ref{EMT} we show that the difference $\langle T_{tt}\rangle_{H}-\langle T_{tt}\rangle_{BD}$ is UV-finite and diverges as $t\rightarrow 0$
\begin{equation}
 \begin{aligned}
  \langle T_{tt}\rangle_{H}-\langle T_{tt}\rangle_{BD} &= -\frac{11}{240\pi^2}\frac{1}{t^4}+O\left(\frac{1}{t^2}\right).
 \end{aligned}
\end{equation}
So we conclude that the hyperbolic vacuum $|\Omega_H\rangle$ has singular properties that are completely analogous to the Minkowski Fulling-Rindler, Milne and de Sitter static vacuum, see also \cite{Eme:2014}. The energy momentum tensor diverges in the limit $t\rightarrow 0$, so infinite energy seems to be required to prepare the state at the (singular) origin. A complete description all the way until $t=0$ is therefore obviously inconsistent, but strictly speaking that does not need to be fatal in a cosmological setting, in the sense that in a stand-alone open universe this might be interpreted as the Big Bang singularity.

Of course, arguably the most natural and well-behaved choice for an initial state on hyperbolic de Sitter sections is the de Sitter invariant Bunch--Davies state, to which we turn next.

\subsection{The Bunch--Davies state in the hyperbolic patch}

Here we will just briefly summarize the results of \cite{Sasaki:1995} and \cite{MalPim:2012}. More details can be found in those papers and in appendix \ref{Modes}. The most important observation is that mode functions of one of the hyperbolic patches (\ref{Hmodes}) do not correspond to regular mode functions on the full (Euclidean) de Sitter space. In \cite{Sasaki:1995} the hyperbolic mode functions are analytically continued to the other hyperbolic patch, allowing them to construct a set of regular mode functions that can cover all of de Sitter space as follows
\begin{equation}\label{continued}
 \begin{aligned}
\chi_p^{(R)}&= \left\{ \begin{matrix}
               P^{ip}_{\nu-\frac{1}{2}}(z) &&\text{for} \ z \in R   \\
               \frac{i \sin(\pi (\nu-\frac{1}{2}))}{\sinh(p\pi)} P^{ip}_{\nu-\frac{1}{2}}(z)
   +\frac{i \sin(\pi(ip+\nu-\frac{1}{2}))e^{}}{\sinh(\pi  p)}\frac{\Gamma[\nu+\frac{1}{2}+ip]}{\Gamma[\nu+\frac{1}{2}-ip]}P^{-ip}_{\nu-\frac{1}{2}}(z) &&\text{for} \ z \in L
               \end{matrix}\right.
 \end{aligned}
\end{equation}
These mode functions do not yet describe the Euclidean or Bunch--Davies vacuum, which can for instance be concluded by the fact that they are not (anti-)symmetric under the transformation $R \leftrightarrow L$. It turns out that the linear combinations $\chi_R\pm\chi_L$ correspond to the proper mode functions associated with the Euclidean or Bunch--Davies vacuum, as was proven by computing the Wightman function \cite{Sasaki:1995}. The (still to be normalized) mode functions are linear combinations of the associated Legendre functions 
\begin{equation}\label{dSmodes}
\begin{aligned}
  \chi_{p,\sigma} &=
 \left\{
 \begin{array}{lr}
  \alpha^{\sigma}_{p,R} P^{ip}_{\nu-\frac{1}{2}}(z)+\beta^{\sigma}_{p,R}
 P^{-ip}_{\nu-\frac{1}{2}}(z) &\text{for} \ x \in R  \\
  \alpha^{\sigma}_{p,L} P^{ip}_{\nu-\frac{1}{2}}(z)+\beta^{\sigma}_{p,L}
 P^{-ip}_{\nu-\frac{1}{2}}(z) &\text{for} \ x \in L 
 \end{array}
 \right.
\end{aligned}\end{equation}
where $\sigma = \pm1$\footnote{$\sigma = \pm 1$ is related to the combination $\chi_P^{(L)}\pm \chi_P^{(R)}$.} and $z = \cosh(t)$; the expressions for the $\alpha$'s and $\beta$'s are given in \eqref{alfabeta}. We stress that the associated Legendre functions $P$ in (\ref{dSmodes}) do not have to be analytically continued any further\footnote{They are constituents of $\chi_{L}$ and $\chi_{R}$, which are already regular everywhere. This is different from \cite{Kanno:2014}.}. The full field expansion, with creation and annihilation operators $\hat{a}_{\sigma plm}$ satisfying $[\hat{a}_{\sigma plm},\hat{a}_{\sigma' p'l'm'}^{\dagger}] = \delta_{\sigma \sigma'}\delta_{ll'} \delta_{mm'} \delta(p-p')$, is given by:
\begin{equation}\label{dSexpansion}
 \begin{aligned}
  \phi(t,r,\Omega) &= \int dp \sum_{\sigma=\pm1}\sum_{l,m}\frac{1}{N_{\chi^{p\sigma}}}\left(\hat{a}_{\sigma plm}\chi_{p,\sigma}(z)Y_{plm}(r,\Omega)+\text{h.c.}\right)
 \end{aligned}
\end{equation}
where $N_{\chi^{p\sigma}}$ is the Klein--Gordon norm consistent with the commutation relations\footnote{Strictly speaking the expression (\ref{dSexpansion}) is incomplete, since we should also include the ``zero mode". For our purposes this will however not affect the results.}. In appendix \ref{KGnorm} we show that the normalization $N_{\chi^{p\sigma}}$ is given by
\begin{equation}\label{dSnorm}
 \begin{aligned}
  N_{\chi^{p\sigma}}^2 &\equiv \langle\chi_{p,\sigma}Y_{plm},\chi_{p, \sigma}Y_{ plm} \rangle_{\text{KG}} \\
  &= \sum_{q = L,R}\left(\alpha^{\sigma}_{p,q}\bar{\alpha}^{\sigma}_{p,q}-\beta^{\sigma}_{p,q}\bar{\beta}^{\sigma}_{p,q}\right)N^2_{P^p}
 \end{aligned}
\end{equation}
where $\bar{\alpha},\bar{\beta}$ denote the complex conjugates of $\alpha,\beta$. We conclude that the Bunch--Davies vacuum state, in a single hyperbolic patch, is defined as\footnote{As before we ignore the supercurvature modes.}
\begin{equation}
 \begin{aligned}
  \hat{a}_{\sigma plm}|\Omega_{BD}\rangle &= 0 &&\forall \, \sigma, p,l,m.
 \end{aligned}
\end{equation}

Now let us describe the relation between the creation and annihilation operators of the modes (\ref{dSmodes}) and the creation and annihilation operators of hyperbolic modes (\ref{Hmodes}). Both field expansions (\ref{dSexpansion}) and (\ref{Hexpansion}) are linear combinations of associated Legendre functions. We can find the relation between the $\hat{a}_{\sigma plm}$ and the $\hat{b}_{qplm}$ ($q = L,R$) by comparing the coefficients  
\begin{equation}\label{relationab}
 \begin{aligned}    
\hat{b}_{q plm} &= \sum_{\sigma=\pm 1}\frac{N_{P^p}}{N_{\chi_{p,\sigma}}}\left(\alpha^{\sigma}_{p,q}\hat{a}_{\sigma plm}+\bar{\beta}^{\sigma}_{p,q}\hat{a}^{\dagger}_{\sigma pl-m}\right).
 \end{aligned}
\end{equation}
Given (\ref{dSnorm},\ref{relationab}), they enforce
\begin{equation}\label{commutation}
 \begin{aligned}
\left.\begin{array}{l}
\left[ \hat{a}_{\sigma plm},\hat{a}^{\dagger}_{\sigma'p'l'm'} \right] =  \delta(p-p')\delta_{\sigma \sigma'}\delta_{mm'}\delta_{ll'} \\ \\
\left[\hat{a}_{\sigma plm},\hat{a}_{\sigma'p'l'm'}\right] =  0
\end{array}\right\}
 \Leftrightarrow \left\{
 \begin{array}{l}
\left[\hat{b}_{qplm},\hat{b}^{\dagger}_{q'p'l'm'}\right]  =  \delta(p-p')\delta_{mm'}\delta_{ll'}\delta_{qq'} \\ \\
\left[\hat{b}_{qplm},\hat{b}_{q'p'l'm'}\right] =  0
\end{array}\right..
 \end{aligned}
\end{equation}

For more details we refer to \cite{Sasaki:1995} and the appendices. The above relationship confirms that the normalizations (\ref{Pnorm}) and (\ref{dSnorm}) are consistent and in particular that the right normalization for the hyperbolic mode functions is given by (\ref{Pnorm}). This will be of importance when comparing the predictions for the power-- and bi--spectrum of the two different states under consideration: the pure hyperbolic vacuum and the Bunch--Davies state (as we will do in section \ref{sec:correlators}). The latter is a mixed state from the point of view of a single hyperbolic patch, due to the entanglement between the two hyperbolic patches in the Bunch--Davies vacuum.

Let us here remind the reader that we would like to compare the predictions for the expectation values of (scalar field) quantum fluctuations in the two different states that were introduced above. A priori different initial states give different predictions for the cosmic microwave background temperature anisotropies and the large scale structure distribution. We should stress that we are technically not considering an actual bubble nucleation event, where more intricate and model-dependent bubble wall physics could lead to additional effects \cite{Sasaki:1996, Gar:1998}, see also \cite{HawHerTur:2000}. Instead, we will work under the assumption that the two states that were introduced capture an essential difference that is generic: the entangled nature of the Bunch--Davies vacuum implies a mixed initial state, whereas the hyperbolic vacuum corresponds to a pure state on a single hyperbolic section. Different (presumably more realistic) states in open universes have been considered in the past \cite{Sasaki:1996, Komatsu:2013}, but these states appear to be pure (and excited) hyperbolic states, so they do not capture any effects related to the mixed nature of the initial state. In the process we hope to clear up some confusion that might have arisen and that could also have consequences for more realistic bubble states that were considered in the past.

We should add that one might anticipate the differences between the two states to only become visible at small hyperbolic momentum $p \lesssim 1$, i.e. scales comparable to the hyperbolic curvature. However, even small curvature suppressed changes in the initial state might be enhanced in the (nonlinear) bi--spectrum, as has been pointed out and analyzed in \cite{Holman:2007, Meerburg:2009} and for a certain generic type of mixed state in \cite{Agullo:2010}. This motivates our particular interest in computing the bi--spectrum, comparing the different states to the planar Bunch--Davies result. But first let us review some general facts regarding correlators in de Sitter space and summarize the results for the two--point functions.

%



\section{Correlators in hyperbolic de Sitter space}\label{sec:correlators}

Making use of the previously established relations between the de Sitter invariant Bunch--Davies vacuum and the hyperbolic vacuum state, we will compute both the Bunch--Davies and the hyperbolic vacuum power--spectra of scalar field quantum fluctuations\footnote{See \cite{Eme:2014} for related work on the response of Unruh detectors.}. The Bunch--Davies result can also be calculated using a reduced density matrix formalism in the hyperbolic patch. Let us from the outset emphasize that within our basic de Sitter set-up, even though the Bunch--Davies state is mixed from the hyperbolic patch perspective, all hyperbolic Bunch--Davies correlators should match the (hyperbolic coordinate transformed) planar Bunch--Davies correlators. As a consequence one can rule out large deviations of Bunch--Davies hyperbolic correlators at late time and large momenta (when the hyperbolic coordinates reduce to planar coordinates) as compared to the planar Bunch--Davies correlators. That leaves the (pure) hyperbolic vacuum as the potentially more interesting state to consider, as far as enhanced initial state effects with respect to the planar Bunch--Davies state are concerned. 

Let us start by pointing out that the field operator $\phi_p$ evaluated on points in the left hyperbolic patch is trivial in the right hyperbolic patch
\begin{equation}
 \begin{aligned}
  \phi_p(x) &= \phi_{L,p}(x)\otimes \mathbf{I}_R &&\text{for} \ x \in \text{L}.
 \end{aligned}
\end{equation}
As a consequence, de Sitter $n$-point functions of fields $\phi_p$ in the Bunch--Davies state, evaluated on points in the left hyperbolic patch, can be calculated either using the full global description or by using a reduced density matrix 
$ \hat{\rho}_L = \text{Tr}_{\mathcal{H}_R}\left\{ |\Omega_{\text{BD}}\rangle\langle\Omega_{\Omega_{BD}}|\right\}$, and their results should agree. This is shown explicitly in appendix \ref{app:reduceddensitymatrixspectrum}. 
By defining the $\hat{b}_{qplm}$ as in (\ref{relationab}), we can write the field operator for arbitrary values of $p,l,m$ as
\begin{equation}
 \begin{aligned}
 \phi_{plm}(x) &= \hat{b}_{Lplm}\frac{1}{N_{P^p}}P^{ip}_{\nu-\frac{1}{2},L}Y_{plm}+\text{h.c.} \\
 &+\hat{b}_{Rplm}\frac{1}{N_{P^p}}P^{ip}_{\nu-\frac{1}{2},R}Y_{plm}+\text{h.c.},
 \end{aligned}
\end{equation}
where 
\begin{equation}
 \begin{aligned}
  P^{ip}_{\nu-\frac{1}{2},L} &= \left\{\begin{array}{ll}
                                 P^{ip}_{\nu-\frac{1}{2}}(t) &\text{for} \ t \in L \\
                                 0 &\text{for} \ t \in R \\
                                \end{array}\right.
 \end{aligned}
\end{equation}
and vice versa for $P^{ip}_{\nu-\frac{1}{2},R}$. Although these functions are not mode functions on a full Cauchy slice covering the de Sitter space, we are allowed to express the field in terms of them. Note that the $\hat{b}_{L}$ and $\hat{b}_{R}$ operators mutually commute. To make explicit that the field operator decomposes in the left and right hyperbolic patches, we write
\begin{equation}\label{explicit}
 \begin{aligned}
 \phi_{plm}(x) &= \left(\hat{b}_{Lplm}\frac{1}{N_{P^p}}P^{ip}_{\nu-\frac{1}{2},L}Y_{plm}+\text{h.c.}\right)\otimes \mathbf{I}_R \\
 &+\mathbf{I}_L\otimes  \left(\hat{b}_{Rplm}\frac{1}{N_{P^p}}P^{ip}_{\nu-\frac{1}{2},R}Y_{plm}+\text{h.c.}\right).
 \end{aligned}
\end{equation}
Note that the expansion of a scalar field in Minkowski spacetime in terms of left and right Rindler wedge modes is similar (see for instance \cite{Parentani:1993}).
The restriction of the operator (\ref{explicit}) to points in the left hyperbolic patch is by definition equal to the full operator evaluated on points in the left hyperbolic patch. 
Note that if the field operator evaluated on points in the left hyperbolic patch would also have support on the right hyperbolic patch, it would not make sense to do a density matrix calculation as done above.  

Clearly therefore Bunch--Davies scalar field correlators should be the same, independent of whether one uses hyperbolic or planar coordinates. Of course, since the de Sitter invariant length is expressed differently in terms of planar or hyperbolic coordinates, the functional dependence of the equal (hyperbolic) time correlators will look different. Since the difference between planar and hyperbolic coordinates vanishes in the late time and large momentum limit, the the planar and the hyperbolic Bunch--Davies correlators match in that limit and small modifications are suppressed in the hyperbolic curvature scale. We conclude that hyperbolic Bunch--Davies correlators can be computed either using a global de Sitter description (for which the Bunch--Davies state is a pure initial state) or by considering a single de Sitter hyperbolic patch (for which the Bunch--Davies state equals a mixed initial state). 

After these important preliminaries let us now proceed by computing the power spectrum of a massless scalar field in a hyperbolic coordinate patch in the hyperbolic vacuum and Bunch--Davies initial state respectively, as a function of the hyperbolic momentum $p$. 

\subsection{Power--spectrum results}
For most of the details we refer to the appendix \ref{app:powerspectramasslessfield}. Here we will just quote the main results. For the two point function in the hyperbolic vacuum state we find 
\begin{equation}
 \begin{aligned}
  \langle \Omega_H|\phi_p\phi_p'|\Omega_H\rangle 
  &= \delta(p-p')\frac{H^2}{\sinh^2(t)}\frac{p}{4\pi^2}\frac{\cosh^2(t)+p^2}{p^2+1}
 \end{aligned}
\end{equation}
where we have used the completeness relation of the eigenfunctions of the hyperbolic Laplacian\footnote{$\sum_{lm}\left|Y_{plm}\right|^2 = \frac{p^2}{2\pi^2}.$} and the commutation relations. At late times $t\rightarrow \infty$ this approaches
\begin{equation}
 \begin{aligned}
  \langle \Omega_H|\phi_p\phi_p'|\Omega_H\rangle &\rightarrow \frac{H^2p}{4\pi^2(p^2+1)}
 \end{aligned}
\end{equation}
and the appropriately normalized power spectrum (at late times) equals 
\begin{equation}\label{psH}
 \begin{aligned}
\Delta^2_{\phi,H}(p) = \frac{H^2}{4\pi^2}\frac{p^2}{p^2+1}.
 \end{aligned}
\end{equation}

The same (hyperbolic coordinate patch) two point function in the Bunch--Davies vacuum is instead found to be equal to 
\begin{equation}\label{2ptdirect2}
 \begin{aligned}
 \langle \Omega_{BD}|\phi_p\phi_{p'}|\Omega_{BD}\rangle &=  \delta(p-p')\frac{H^2}{\sinh^2(t)}\frac{p}{4\pi^2}\frac{\cosh^2(t)+p^2}{p^2+1}\coth(\pi p).
 \end{aligned}
\end{equation}
This result can either be obtained from a direct calculation using the global Bunch--Davies vacuum construction and restricting to one of the hyperbolic coordinate patches \cite{Sasaki:1995}, or from a (mixed) density matrix calculation in a single hyperbolic coordinate patch, using the explicit expression for the density matrix as reported in \cite{MalPim:2012}, as we confirm in appendix \ref{app:reduceddensitymatrixspectrum}.
As alluded to earlier, the reason for this expression to not exactly reproduce the scale-invariant planar coordinate result for the scalar field power spectrum in the Bunch--Davies vacuum is that different coordinates are used. As the hyperbolic and planar coordinates are the same at late times and for small distances, the late-time power spectra at large momentum should be the same as well. At late times $t\rightarrow \infty$ we find
\begin{equation}
 \begin{aligned} 
  \langle \Omega_{BD}|\phi_p\phi_{p'}|\Omega_{BD}\rangle &\rightarrow \frac{H^2p}{4\pi^2(p^2+1)}\coth(\pi p).
 \end{aligned}
\end{equation}
Correspondingly, the power spectrum (at late times) is given by
\begin{equation}\label{psBD}
 \begin{aligned}
\Delta^2_{\phi,BD}(p) =\frac{H^2}{4\pi^2}\frac{p^2}{p^2+1}\coth(\pi p).
 \end{aligned}
\end{equation}

\begin{figure}[ht]
\centering
\begin{minipage}[t]{\textwidth}
\centering
\includegraphics[width=0.9\textwidth]{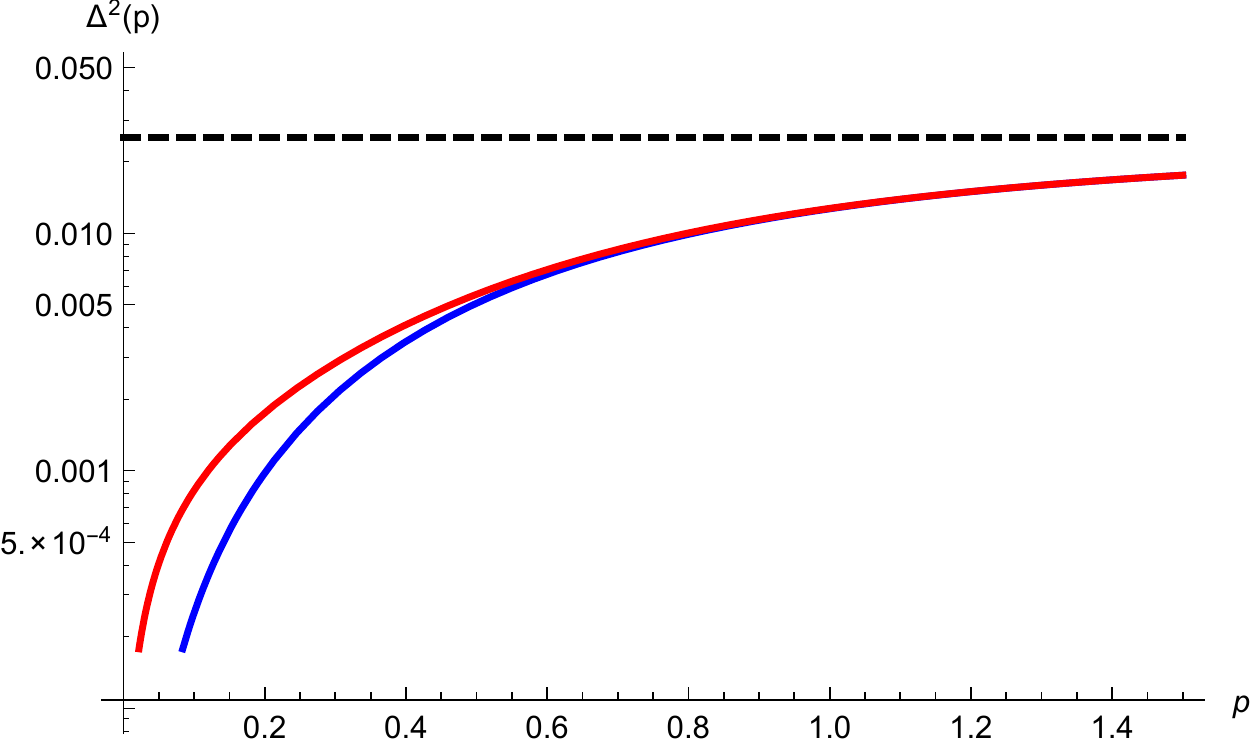}\caption{The power spectra (logarithmic scale) for the hyperbolic vacuum (blue) and the Bunch--Davies vacuum (red), as function of the hyperbolic momentum $p$ with $H=1$. The dashed line indicates the scale--invariant planar Bunch--Davies result $\frac{H^2}{4 \pi^2}$. }\label{fig1}
\end{minipage}
\end{figure}  

Looking at these power spectra we indeed find that for $p \gg 1$, when $\coth\pi p\approx 1$ and $p^2+1 \approx p^2$, the hyperbolic vacuum as well as the hyperbolic Bunch--Davies result matches the standard scale invariant planar Bunch--Davies result $\frac{H^2}{4\pi^2}$. They only start to differ from each other and the standard planar Bunch--Davies expression for sufficiently small momenta $p \lesssim 1$ (see figure (\ref{fig1})). Note that although the corresponding wavelengths are expected to lie far outside our observable window, given the fact that they correspond to length scales longer or comparable to the hyperbolic curvature scale, for both hyperbolic states the power is suppressed as compared to the standard planar Bunch--Davies result. As these departures from the standard planar result become evident, one should keep in mind that the difference found in the hyperbolic Bunch--Davies result can be attributed to a coordinate change, whereas (part of) the change in the hyperbolic vacuum power spectrum is related to an initial state modification. Although this difference might seem unimportant at this point, when considering the bi--spectrum in the next section it is a relevant distinction, since the bi--spectrum has been found to be particularly sensitive to changes in the initial state.  

Finally let us also remark that the bubble state put forward in \cite{Sasaki:1996} and more recently used in \cite{Komatsu:2013} does resemble the Bunch--Davies state in the sense that the expectation value of the number operator on hyperbolic sections agrees (giving thermal occupation numbers in terms of co-moving momentum), but in one aspect is crucially different due to the fact that it has been constructed as a Bogoliubov transformation of the hyperbolic vacuum. As a consequence the state necessarily corresponds to a pure state (and inherits the singular properties of the hyperbolic vacuum) and it explains why their result for the power spectrum does not agree with our hyperbolic Bunch--Davies result that necessarily corresponds to a mixed state.

We conclude that, independent of the particular initial state under consideration, any power--spectrum signatures of an open inflationary universe are confined to the curvature scale, which has to be several orders of magnitude larger than the largest observable length scale in the universe. Although the initial hyperbolic state is mixed when assuming a (globally defined) planar Bunch--Davies state, the power--spectrum results in this admittedly basic set-up in which all bubble wall physics is ignored, do not show large deviations, as should be expected. In fact, this statement is true for general hyperbolic $n$-point correlators in the Bunch--Davies vacuum. Potentially enhanced bi--spectrum results due to initial state excitations, as compared to the standard planar Bunch--Davies result, can therefore only be expected assuming the pure hyperbolic vacuum as the initial state in the hyperbolic patch.


\subsection{The bi--spectrum in the hyperbolic vacuum}

Let us next consider the bi--spectrum of scalar density perturbations in the hyperbolic vacuum, which can be viewed (for any finite boost parameter $\gamma$) as some sort of excited initial state with respect to the standard planar Bunch--Davies vacuum. We will mainly be interested in the so--called squeezed limit, for which previous work uncovered enhanced results for excited (planar) Bunch--Davies intial states \cite{Holman:2007, Meerburg:2009, Agullo:2010, Holman:2012}. Moreover, as before it should be reasonable to work in the sub--curvature approximation $p \gg 1 $, for which the hyperbolic momenta $p_i$ are approximately equal to the standard (flat) wavenumbers $k_i$ up to curvature suppressed corrections.  

To compute the hyperbolic bi--spectrum we need the action to third order in the scalar density perturbation $\varphi$ and the hyperbolic curvature introduces some new ingredients as compared to the planar calculation \cite{Maldacena:2002}, which have been carefully dealt with in \cite{Komatsu:2013}. The state of interest in \cite{Komatsu:2013} is however the bubble state constructed first in \cite{Sasaki:1996}, which is related to the hyperbolic vacuum by means of a Bogoliubov transformation and crucially differs from the Bunch--Davies vacuum. The bubble state can be viewed as a (well-motivated and first principles derived) initial state modification with respect to the hyperbolic vacuum, explaining their interest in trying to identify enhanced features in the bi--spectrum. Note that their computation is not applicable to initial state modifications of the (planar) Bunch--Davies vacuum. Such an interpretation would only be valid in the infinite boost limit, when the hyperbolic vacuum reduces to the planar Bunch--Davies vacuum. As we explained, in our pure de Sitter set-up it is the hyperbolic vacuum initial state that (for finite boost parameter $\gamma$) corresponds to an excited state as compared to the planar Bunch--Davies vacuum and could therefore potentially display interesting bi--spectrum enhancements. Instead the hyperbolic Bunch--Davies bi--spectrum equals the planar Bunch--Davies bi--spectrum and all apparent changes can be related to the coordinate change from planar to hyperbolic. So here we will be interested in computing the results for bi--spectrum in the hyperbolic vacuum, which can fortunately be extracted straightforwardly from the results in \cite{Komatsu:2013}. 

So let us first briefly review the basic results reported in \cite{Komatsu:2013} and then apply them to our case of interest. We will be interested in a massless minimally coupled scalar field, for which the positive and negative frequency modes defined in a single hyperbolic patch, can be nicely expressed as
\begin{eqnarray}\label{eq:modefunctions}
u_p (\eta) & = & H \frac{\cosh \eta + ip \sinh \eta}{\sqrt{2p(1+p^2)}}e^{-ip\eta}, \nonumber \\
v_p (\eta) & = & H \frac{\cosh \eta - ip \sinh \eta}{\sqrt{2p(1+p^2)}}e^{ip\eta}.
\end{eqnarray}
The perturbed metric in one of the hyperbolic patches, is written in the ADM formalism as
\begin{eqnarray}
ds^2 = - N^2 dt^2 + h_{ij}\left( dx^i + N^i dt \right)\left( dx^j + N^j dt \right),
\end{eqnarray}
where, as usual, $N$ is the lapse function, $N^i$ is the shift and $h_{ij} = a^2 (t) e^{2 \zeta} \gamma _{ij} $ is the spatial metric, with curvature perturbation $\zeta$. We will be interested in gravitationally induced  nonlinearities on the scalar density perturbation, requiring that we need to introduce (generic) slow-roll evolution of the background scalar $\phi(t)$ in order to couple the scalar inflaton field to the scalar density perturbation. Now, plugging the above metric into the action for the scalar degree of freedom (assuming slow-roll evolution) and solving the constraint equations order by order, one can obtain the quadratic and cubic (and higher) action. The quadratic action for scalar perturbations, in the flat gauge $\zeta = 0$, to leading order in the slow roll parameters is
\begin{eqnarray}
S^{(2)} = \int dt d^3x a(t)^3 \sqrt{\gamma} \left( \frac{1}{2} \dot{\varphi}^2 - \frac{1}{2 a(t)^2} \partial _i \varphi \partial ^i \varphi \right),
\end{eqnarray}
where $\partial _i$ is the covariant derivative with respect to $\gamma _{ij}$. Up to this point this should all be familiar, so let us now turn to the cubic action on a hyperbolic patch. As explained in \cite{Komatsu:2013} the dominant term in the third order Lagrangian is found to be 
%
%
%
%
%
\begin{eqnarray}\label{eq:dominantterm}
\mathcal{L} ^{(3)} & = & - \sqrt{\gamma} a^{5} \dot{\phi} \left( \left( \partial ^{2} - 3 \right)^{-1}\dot{\varphi}_c \right) \dot{\varphi}_c^{2}, 
\end{eqnarray}
The new ingredient due to the hyperbolic curvature in this action is the $-3$ term. The $\varphi _c$ is (as usual) the redefined field $\left( \varphi \rightarrow \varphi _c \right)$, defined as
\begin{eqnarray}
\varphi = \varphi _c + \frac{\dot{\phi}}{4 (\dot{a}/a)} \left[ \left( \partial ^2 - 3 \right)^{-1} \partial _i\varphi _c \partial ^i \varphi _c \right],
\end{eqnarray}
which does not affect the quadratic action and has removed terms in the cubic action that are proportional to the equations of motion. The bi--spectra of $\varphi$ and $\varphi _c$ are then related as follows
\begin{eqnarray}
\left \langle \varphi (x_1) \varphi (x_2) \varphi (x_3) \right \rangle & = & \left \langle \varphi _c (x_1) \varphi _c (x_2) \varphi _c (x_3) \right \rangle \nonumber \\
         & + &  \frac{\dot{\phi}}{4(\dot{a}/a)} \left[ \left( \partial ^{2} - 3 \right)^{-1} \left \langle \partial _i \varphi _c(x_1) \varphi _c (x_2) \right \rangle \left \langle \partial ^i \varphi _c(x_1) \varphi _c (x_3) \right \rangle + \text{permutations} \right] \nonumber \\
\end{eqnarray}
In harmonic space, introducing the geometrical factor 
\begin{equation}
\int d^3x \sqrt{\gamma} Y_{p_1 l_1 m_1}(x) Y_{p_2 l_2 m_2}(x) Y_{p_3 l_3 m_3}(x) \equiv {\cal F}^{l_1 l_2 l_3}_{p_1 p_2 p_3} {\cal G}^{m_1 m_2 m_3}_{l_1 l_2 l_3} \, , 
\end{equation}
the bi--spectrum $B(p_1, p_2, p_3)$ is defined as 
\begin{equation}
\left \langle \varphi_{p_1 l_1 m_1} \varphi_{p_2 l_2 m_2} \varphi_{p_3 l_3 m_3} \right \rangle = B(p_1, p_2, p_3) \, {\cal F}^{l_1 l_2 l_3}_{p_1 p_2 p_3} {\cal G}^{m_1 m_2 m_3}_{l_1 l_2 l_3} \, .
\end{equation}
Of the two bi--spectrum contributions due to the field redefinition, the first term is then computed using eq.~(\ref{eq:dominantterm}) in the in--in formalism, where we already assumed the sub-curvature limit allowing one to replace the momentum $p$ with the planar momentum $k$ 
\begin{eqnarray}\label{eq:bispectrum}
B(k_1,k_2,k_3) & = & 2 \text{Re} \left[ i v_{k_1}(0) v_{k_2}(0) v_{k_3}(0) \left( \int_{-\infty}^{0} d \eta \frac{2 a^6 \dot{\phi}}{k_1 ^2 + 4} \dot{u}_{k_1}(\eta) \dot{u}_{k_2}(\eta) \dot{u}_{k_3}(\eta) \right) + 1 \leftrightarrow 2 + 1 \leftrightarrow 3 \right]. \nonumber \\ 
\end{eqnarray}
In a hyperbolic (quasi) de Sitter space the integral over conformal time is naturally divided into two eras: from $-\infty$ to $-1$, where the curvature term is dominant and from $-1$ to $0$ corresponding to standard inflationary expansion. Assuming slow-roll evolution in both eras, the expressions for $\dot{\phi}$ in terms of the slow-roll parameter $\epsilon$ change when transitioning from the curvature-dominated era into the inflationary era.
The mode functions defining the hyperbolic vacuum $| \Omega_H \rangle$ in the sub--curvature approximation $p \gg 1 $, in the different eras, read as follows
\begin{eqnarray}
u_k(\eta) &\simeq& -\frac{i}{a(\eta)\sqrt{2k}} e^{-ik\eta} \quad (-\infty < \eta \lesssim -1) \, \nonumber \\
u_k(\eta) &\simeq& -\frac{H}{\sqrt{2k^3}} (1+ik\eta) e^{-ik\eta} \quad (-1 \lesssim \eta < 0) \, \nonumber \\
\dot{u}_k (\eta) &=& \frac{\partial u_k}{\partial \eta} \frac{1}{a(\eta)} \simeq  \frac{1}{a^{2}(\eta)} \sqrt{\frac{k}{2}} e^{-ik \eta}, \nonumber \\
v_{k}(0) & \simeq & \frac{H}{\sqrt{2k^3}}.
\end{eqnarray}
Using these expressions to do the integral in (\ref{eq:bispectrum})and ignoring effects due to the discontinuity in the transition between eras, one obtains \cite{Komatsu:2013}
\begin{eqnarray}
B(k_1, k_2, k_3)  & = & \frac{2 \sqrt{2\epsilon}H^4}{4 k_1 k_2 k_3\left( k_1 + k_2 + k_3 \right)} \left(  \frac{1}{ k_1 ^2} + \frac{1}{k_2 ^2} + \frac{1}{ k_3 ^2} \right).
\label{hyperbolic-bispectrum}
\end{eqnarray}
This result agrees with the standard planar Bunch--Davies result in single field inflation. For completion the additional contribution from the field redefinition equals 
\begin{eqnarray}
B^{\text{redef}} (k_1 , k_2 , k_3) & = & \frac{\dot{\phi}}{4H} \frac{\mathbf{k_1} \cdot \mathbf{k_2}}{k_3 ^2 + 4} \frac{H ^2}{2 k_1 ^3} \frac{H ^2}{2 k_2 ^3} + \text{permutations} \, ,
\end{eqnarray}
but as shown in \cite{Komatsu:2013} the contribution from this term is subdominant in the squeezed limit $k_3 \ll k_1 \simeq k_2 = k$ and therefore can be ignored when looking for other enhanced contributions in the squeezed limit, as expected for initial state modifications. The enhancement of initial state modifications in the squeezed limit, as compared to the hyperbolic vacuum, was indeed observed after adding the specific negative frequency term to the mode functions that describes the bubble initial state \cite{Komatsu:2013}. However the Bogoliubov coefficient describing this excited bubble state is exponentially suppressed in the hyperbolic momentum over the curvature scale (which is normalized to one) $e^{-\pi p}$, meaning that these effects for the observable modes in the sky are exponentially suppressed and undetectable. 

After having reviewed the hyperbolic bi--spectrum calculation in \cite{Komatsu:2013}, we will now connect these results to our case of interest. To start out with, the quoted result for the bi--spectrum (\ref{hyperbolic-bispectrum}) is of course the hyperbolic vacuum bi--spectrum in the late time, sub-curvature limit, which agrees with the standard planar Bunch--Davies result. Clearly then, no (modified initial state) enhancement with respect to the planar Bunch--Davies result is found in the late time and large momentum limit. The squeezed enhancement that was revealed in \cite{Komatsu:2013} for the excited bubble state, albeit exponentially suppressed in momentum, is with respect to the hyperbolic vacuum and can only be interpreted as enhancement with respect to the planar Bunch--Davies vacuum in the late time and large momentum limit. As we noted, the late time and large momentum limit effectively corresponds to the action of the infinite boost in the embedding space for which the hyperbolic vacuum indeed matches the planar Bunch--Davies vacuum state, explaining the result. The excited bubble vacuum, obtained through a Bogoliubov rotation from the hyperbolic vacuum and therefore corresponding to a pure state, will also not agree with the hyperbolic Bunch--Davies result. Instead the hyperbolic bi--spectrum in the Bunch--Davies state has to match the planar Bunch--Davies result (written in hyperbolic coordinates), obviously excluding a (squeezed) enhancement. 

Our first conclusion is therefore that the bi--spectrum in the hyperbolic vacuum is clearly (and unsurprisingly) not enhanced with respect to the planar Bunch--Davies result in the late time and sub-curvature limit. In fact it is straightforward to (perturbatively) extend this conclusion beyond the strict sub-curvature or large momentum limit by analyzing the leading correction. The first correction in the large momentum approximation is obtained from an asymptotic expansion of the hyperbolic mode function at late times 
\begin{equation}
u_p (\eta \ll 1, p \gg 1) \approx \frac{H}{\sqrt{2p^3}} \, (1 + ip \eta) \, e^{-ip\eta} \, \left( 1-\frac{1}{2p^2} + {\cal O}(1/p^4) \right) \, . \\
\label{pexpansion}
\end{equation}
Noting the last term (in brackets) one observes that the leading large momentum correction will give additional contributions suppressed in the large momentum limit with at least one factor of $1/p^2$\footnote{Re-installing the hyperbolic curvature scale, which was set to one, this term would be explicitly dimensionless and read $k_c/p^2$.}, but leaving the relative momentum dependence in tact. As a consequence enhancements in the squeezed momentum limit as compared to the planar Bunch--Davies result are excluded, as that would typically require additional terms featuring negative frequencies. It should be clear that corrections suppressed as $1/p^2$, without additional enhancements in particular momentum configurations, does not constitute an interesting non-Gaussian signature of an open inflationary universe. 

We conclude that, unfortunately, the bi--spectrum in the hyperbolic vacuum does not produce interesting enhancements that could be searched for. This conclusion is in fact corroborated by effectively constructing the hyperbolic vacuum as an excited state on top of the (global) planar Bunch--Davies vacuum (see appendix \ref{app:Hvacuum}). Effectively one then discovers that the relevant Bogoliubov coefficients are suppressed exponentially in momentum, ensuring that these effects will not be observable. As emphasized, the same is true for the bi--spectrum in the hyperbolic Bunch--Davies initial state. So even though the hyperbolic vacuum can be thought of as an excited state with respect to standard planar Bunch--Davies vacuum, it is not of a type that gives rise to large (enhanced) corrections in the bi--spectrum as compared to the standard planar Bunch--Davies result.

%
%
%
%
%
%


\section{Conclusions}

Before summarizing our results, let us remind the reader once more that our motivation was to carefully study the relation between the hyperbolic and planar coordinate patches and their corresponding states in (mostly) pure de Sitter space. We believe these results to be of interest, and partially applicable, in the context of de Sitter false vacuum decay, but it is also clear that in that case a more complete analysis should include (model-dependent) wall physics that will affect the details. Instead we concentrated on a general and qualitative difference between two examples of initial states on a hyperbolic section of de Sitter space: the pure hyperbolic vacuum and the (entangled) de Sitter invariant Bunch--Davies state, corresponding to a mixed state. We first of all noted that the pure hyperbolic vacuum is formally inconsistent, due to the energy momentum tensor becoming singular at the null boundary of the hyperbolic section. This issue should plague all pure hyperbolic states, in particular also the bubble state introduced in \cite{Sasaki:1996}, which is obtained from the hyperbolic vacuum by a (unitary) Bogoliubov transformation. 

The Bunch--Davies vacuum is qualitatively different, giving rise to a mixed state on a single hyperbolic section. The fact that the Bunch--Davies state is mixed on a hyperbolic section could a priori imply interesting non-Gaussian signatures \cite{Kanno:2014} as compared to the bi--spectrum in the hyperbolic vacuum. But as we pointed out all planar $n$-point correlators in the Bunch--Davies state, with all points in the hyperbolic section, should equal the hyperbolic $n$-point Bunch--Davies correlators. Any differences are generated by the coordinate change from planar to hyperbolic, instead of arising due to changes in the initial state. So even though the Bunch--Davies state is a mixed state on a hyperbolic section, certainly implying different results as compared to the hyperbolic vacuum, it will reproduce all planar Bunch--Davies correlators, up to coordinate changes. Since the hyperbolic and planar coordinates coincide in the late time and large momentum (sub-curvature) limit all effects due to the coordinate change should disappear and the leading corrections will not be enhanced. 

We stressed that instead the hyperbolic vacuum should generically be considered as an excited state as compared to the Bunch--Davies vacuum. As a consequence that should be the state of interest for computing the bi--spectrum and look for enhanced features. Using the hyperbolic coordinate embedding we explicitly constructed a family of hyperbolic solutions that reduces to the planar coordinates in the infinite boost limit, as such providing a limiting relation between the hyperbolic vacuum and the planar Bunch--Davies vacuum. As a corollary we also argued that the hyperbolic vacuum can be mapped to a specific static vacuum, implying that the static vacuum should also reduce to (a sector of) the Bunch--Davies state in the infinite boost limit, as was first noted in \cite{GreParScha:2006}. Again, this limiting behavior implies that in the late time and large momentum limit, the bi--spectrum results for the hyperbolic vacuum should agree with the standard planar Bunch--Davies result and using the results of \cite{Komatsu:2013} this was indeed confirmed. Looking at the leading correction in the large momentum expansion, we verified that no enhancement in particular momentum configurations is generated and that the corrections are at least suppressed as $1/p^2$. These type of curvature suppressed Non-Gaussian corrections will clearly be impossible to detect. So unfortunately, on the basis of our analysis here, we should conclude that no detectable signals of an open inflationary universe in the fluctuation statistics is expected on small sub-curvature scales. 

To summarize the two hyperbolic states introduced, one of them mixed (Bunch--Davies) and the other one pure (hyperbolic vacuum), make almost identical predictions in the late time sub-curvature limit. In fact, in the infinite boost limit the states become formally identical to the planar Bunch--Davies vacuum. For the mixed Bunch--Davies state this seems to imply that the density matrix $\rho_{BD}$ should depend on the boost parameter $\gamma$. Correspondingly, the associated von Neumann entropy $\text{Tr} (-\rho_{\text{BD}} \, \ln \rho_{\text{BD}})$ of the mixed Bunch--Davies state on the hyperbolic section should depend on the boost parameter $\gamma$ to ensure that the entropy vanishes in the infinite boost limit. This density matrix was computed in \cite{MalPim:2012} and it would be of interest to consider the generalization for non-zero boost parameter $\gamma$. Although one might think the density matrix and corresponding entropy to be boost invariant, this is not entirely obvious and the above observation does indeed suggest it might not be, perhaps in some subtle (singular) way. The dependence on the boost parameter should be such that it is invariant under $\gamma \rightarrow -\gamma$, effectively interchanging the two hyperbolic sections. Since the boost dependence can be implemented through a simple rescaling on the left hyperbolic momenta (and a time shift) \ref{fixedcoord} and the inverse rescaling on the right hyperbolic momenta (and time shift) it should be possible to trace the boost dependence of the Bunch--Davies state in terms of the left and right hyperbolic modes. It should then be straightforward to construct the corresponding density matrix and confirm that the density matrix and corresponding entropy become trivial in the infinite boost limit. 

In more realistic scenarios, trying to incorporate the bubble nucleation dynamics, an initial state has been proposed that can be constructed by performing a unitary Bogoliubov transformation on the hyperbolic vacuum state \cite{Sasaki:1996}. As a consequence this state appears to be pure, which at first sight contradicts the general expectation that the initial state inside a bubble, after false vacuum decay, should be mixed. Although the pure bubble state is in certain aspects similar to the Bunch--Davies vacuum, the correlation functions will be different as compared to the Bunch--Davies hyperbolic correlation functions. In addition, as in the hyperbolic vacuum, the expectation value of the stress tensor in this pure hyperbolic bubble state should become singular as one approaches the bubble wall. It might be of interest to revisit the original construction and see how it can be adjusted to construct a mixed hyperbolic initial state instead. We hope to address this and some of the other remaining questions in future work.

\acknowledgments

We thank I-Sheng Yang and Ben Freivogel for useful discussions. This work is part of the Delta ITP consortium, a program of the Netherlands Organisation for Scientific Research (NWO) that is funded by the Dutch Ministry of Education, Culture and Science (OCW). This work is also supported in part by the Foundation for Fundamental Research (FOM), which is part of NWO.


\begin{appendix} 
\section{Mode functions}\label{Modes}
\subsection{Solutions to the hyperbolic equation of motion}
The metric for both the left and right hyperbolic patch is given by:
\begin{equation}
 \begin{aligned}
  ds^2 &= \frac{1}{H^2}\left(-dt^2+\sinh^2t\left(dr^2+\sinh^2r \ d\Omega_2^2\right)\right)
 \end{aligned}
\end{equation}  
where the coordinates $t,r,\phi,\theta$ are dimensionless and $c=1$. The action for a massive non-interacting minimally coupled scalar field $\phi$ is given by:
\begin{equation}
 \begin{aligned}
  S &= -\frac{1}{2}\int \sqrt{-g} \ d^4x \left(g^{\mu\nu}\partial_{\mu}\phi\partial_{\nu}\phi+m^2\phi^2\right). 
 \end{aligned}
\end{equation}
The action of a conformally coupled scalar field can be written in the Einstein frame with effective mass $m^2 = 2H^2$. The equation of motion is given by
\begin{equation}
 \begin{aligned}
0 &= \left({\frac{1}{\sinh^3(t)}}\frac{\partial}{\partial t}\sinh^3(t)
\frac{\partial}{\partial t}
-\frac{1}{\sinh^2(t)} \nabla_{\mathcal{H}^3}^2+\frac{m^2}{H^2}\right)\phi \\
&= \left({\frac{1}{\sinh^3(t)}}\frac{\partial}{\partial t}\sinh^3(t)
\frac{\partial}{\partial t}
-\frac{1}{\sinh^2(t)} \nabla_{\mathcal{H}^3}^2+\frac{9}{4}-\nu^2\right)\phi
\end{aligned}
\end{equation}
where $\nu$ is defined as $\nu=\sqrt{\frac{9}{4}-\frac{m^2}{H^2}}$ and $\nabla_{\mathcal{H}^3}^2$ is the Laplacian on the hyperboloid $\mathcal{H}^3$. We will use $\nu' = \nu -\frac{1}{2}$, consistent with \cite{Sasaki:1995}, such that $\nu' = 1$ corresponds to the massless minimally coupled case and $\nu' = 0$ corresponds to the massless conformally coupled case for which the effective mass is $m^2 = 2H^2$. 

The eigenfunctions $Y_{plm}$ of the Laplacian $\nabla_{\mathcal{H}^3}^2$ on the hyperboloid $\mathcal{H}^3$, that are regular in $r=0$, are given by \cite{Sasaki:1995}:
\begin{eqnarray}
-\nabla_{\mathcal{H}^3}^2Y_{plm} &=& (p^2+1)Y_{plm} \nonumber\\
Y_{plm}(r,\Omega)&=&f_{pl}(r)Y_{lm}(\Omega)
\nonumber\\
 f_{pl}(r)&=&
\frac{\Gamma(ip+l+1)}{\Gamma(ip+1)}\frac{p}{\sqrt{\sinh r}}
   P^{-l-1/2}_{ip-1/2}(\cosh r)
\nonumber\\
&=&
(-1)^l\sqrt{\frac{2}{\pi}}\,\frac{\Gamma(-ip+1)}{\Gamma(-ip+l+1)}\,
  \sinh^lr \frac{d^l}{d(\cosh r)^l}
    \left(\frac{\sin pr}{\sinh r}\right)
\label{fpl}
\end{eqnarray}
where $Y_{lm}(\Omega)$ is the normalized
spherical harmonic function on the unit two-sphere,
$\Gamma(z)$ is the Gamma function and $P^{\nu}_{\mu}(z)$ is
the associated Legendre function of the first kind \cite{Magnus}.

The mode functions that correspond to the natural hyperbolic vacuum are given by:
\begin{equation}
 \begin{aligned}
  \left\{\begin{array}{ll}
          \frac{H}{\sinh t}P^{ip}_{\nu'}\left(\cosh t\right) &\text{positive energy modes} \\
          \frac{H}{\sinh t}P^{-ip}_{\nu'}\left(\cosh t\right)
          &\text{negative energy modes}
         \end{array}
: p\geq 0\right\}.
 \end{aligned}
\end{equation}

Mode functions on a Cauchy slice of de Sitter, must be regular and consist of linear combinations of the hyperbolic mode functions in the left and right hyperbolic patches \cite{Sasaki:1995}. The mode functions that correspond to the Bunch--Davies state are given in \cite{Sasaki:1995}:
\begin{equation}\label{tanakaexpans}
\begin{aligned}
 \chi_{p,\sigma} &=
 \left\{
 \begin{array}{ll}
  \displaystyle
 \left(
 {e^{\pi p}-\sigma e^{-i\pi\nu'}\over \Gamma(\nu'+ip +1)}
 P^{ip}_{\nu'}(z)-
 {e^{-\pi p}-\sigma e^{-i\pi\nu'}\over \Gamma(\nu'-ip +1)}
 P^{-ip}_{\nu'}(z)\right) &\text{for} \ x \in R 
 \\
 \displaystyle
  \left(
  {\sigma e^{\pi p}- e^{-i\pi\nu'}\over \Gamma(\nu'+ip +1)}
 P^{ip}_{\nu'}(z)-
  {\sigma e^{-\pi p}- e^{-i\pi\nu'}\over \Gamma(\nu'-ip +1)}
 P^{-ip}_{\nu'}(z)\right) &\text{for} \ x \in L
 \end{array}
 \right. \\
 &= \left\{
 \begin{array}{lr}
  \alpha^{\sigma}_{p,R} P^{ip}_{\nu'}(z)+\beta^{\sigma}_{p,R}
 P^{-ip}_{\nu'}(z) &\text{for} \ x \in R  \\
  \alpha^{\sigma}_{p,L} P^{ip}_{\nu'}(z)+\beta^{\sigma}_{p,L}
 P^{-ip}_{\nu'}(z) &\text{for} \ x \in L 
 \end{array}
 \right.
 \end{aligned}
\end{equation}
where $z = \cosh t$ and the constants $\alpha^{\sigma}_{p,q}$ and $\beta^{\sigma}_{p,q}$ are defined as:
\begin{equation}\label{alfabeta}
 \begin{aligned}
  \begin{array}{ll}
   \alpha^{\sigma}_{p,L} = \sigma {e^{\pi p}-\sigma e^{-i\pi\nu'}\over \Gamma(\nu'+ip +1)} & \alpha^{\sigma}_{p,R} = {e^{\pi p}-\sigma e^{-i\pi\nu'}\over \Gamma(\nu'+ip +1)} \\
   \beta^{\sigma}_{p,L} = -\sigma
 {e^{-\pi p}-\sigma e^{-i\pi\nu'}\over \Gamma(\nu'-ip +1)} & \beta^{\sigma}_{p,R} = -
 {e^{-\pi p}-\sigma e^{-i\pi\nu'}\over \Gamma(\nu'-ip +1)} 
  \end{array}. 
 \end{aligned}
\end{equation}
These mode functions must be normalized through the Klein--Gordon normalization (see section \ref{KGnorm}).

\subsection{Klein--Gordon normalization}\label{KGnorm}
\subsubsection*{Hyperbolic modes}
We normalize the hyperbolic modes on the hyperbolic patch using the variable $z = \cosh t$ and using the orthonormality of the $Y_{plm}$:
\begin{equation}\label{NP}
 \begin{aligned}
N_{P^p}^2 &\equiv \langle\phi_{plm},\phi_{plm} \rangle_{\text{KG}} \\ 
&= i \int_{\Sigma}d\Sigma^{\mu}\left(\phi_{plm}\partial_{\mu} \phi_{plm}^*-\phi_{plm}^*\partial_{\mu} \phi_{plm}\right) \\
&= i \sinh^3 t \left(\frac{P^{ip}_{\nu'}(\cosh t)}{\sinh t}\partial_t\left( \frac{P^{-ip}_{\nu'}(\cosh t)}{\sinh t}\right)-\frac{P^{-ip}_{\nu'}(\cosh t)}{\sinh t}\partial_t \left(\frac{P^{ip}_{\nu'}(\cosh t)}{\sinh t }\right)\right) \\
&= i (z^2-1)\left(P^{ip}_{\nu'}(z)\partial_z P^{-ip}_{\nu'}(z)-P^{-ip}_{\nu'}(z)\partial_z P^{ip}_{\nu'}(z)\right).
 \end{aligned}
\end{equation}
For the minimally coupled massless case $\nu' = 1$ we have:
\begin{equation}\label{pnorm}
 \begin{aligned}
N^2_{P^p} &= i (z^2-1) P^{ip}_1(z)P^{-ip}_1(z)\times \left(\frac{1}{z-ip}+\frac{ip}{2}\frac{1}{1+z}+\frac{ip}{2}\frac{1}{1-z}- \frac{1}{z+ip}+\frac{ip}{2}\frac{1}{1+z}+\frac{ip}{2}\frac{1}{1-z} \right) \\
&= \frac{2p}{\left|\Gamma[1+ip]\right|^2} \\
&= \frac{2\sinh(\pi p)}{\pi}.
 \end{aligned}
\end{equation}
In fact, for $\nu' \neq 1$ this normalization is also valid. In \cite{Sasaki:1995} it is shown that one can expand the mode functions in the $t\rightarrow 0$ regime:
\begin{equation}
 \begin{aligned}
  \frac{1}{\sinh t}P^{ip}_{\nu'}(\cosh t) &\approx  \frac{2^{ip}}{\Gamma[1-ip]}t^{ip-1}.
 \end{aligned}
\end{equation}
Using this expansion in (\ref{NP}) also results into the normalization (\ref{pnorm}). This normalization is valid for any $t$ by the properties of the Klein--Gordon normalization. 

\subsubsection*{Bunch--Davies modes}
The Bunch--Davies modes are given in terms of linear combinations of the hyperbolic modes in (\ref{tanakaexpans}). Schematically we have (using the orthogonality of the hyperbolic mode functions):
\begin{equation}\label{Nchi}
 \begin{aligned}
  N^2_{\chi^{\sigma,p}} &= \langle \chi_{\sigma, p} ,\chi_{\sigma, p} \rangle \\
  &= \sum_{q=L,R}  \sum_{q'=L,R} \langle\left(\alpha^{\sigma}_{p,q}P^{p,q}+\beta^{\sigma}_{p,q}\bar{P}^{p,q}\right),\left(\alpha^{\sigma}_{p,q'}P^{p,q'}+\beta^{\sigma}_{p,q'}\bar{P}^{p,q'}\right)\rangle_{KG} \\
  &= \sum_{q,q'=L,R}\left(\begin{array}{l} \alpha^{\sigma}_{p,q}\bar{\alpha}^{\sigma}_{p,q'}\langle P^{p,q},P^{p,q'}\rangle_{KG} + \alpha^{\sigma}_{p,q}\bar{\beta}^{\sigma}_{p,q'}\langle P^{p,q},\bar{P}^{p,q'}\rangle_{KG} \\ +\bar{\alpha}^{\sigma}_{p,q'}\beta^{\sigma}_{p,q}\langle \bar{P}^{p,q},P^{p,q'}\rangle_{KG}+\beta^{\sigma}_{p,q}\bar{\beta}^{\sigma}_{p,q'}\langle \bar{P}^{p,q},\bar{P}^{p,q'}\rangle_{KG}
\end{array}\right) \\
&= N^2_{P^p}\sum_{q=L,R}\left(\alpha^{\sigma}_{p,q}\bar{\alpha}^{\sigma}_{p,q}-\beta^{\sigma}_{p,q}\bar{\beta}^{\sigma}_{p,q}\right),
 \end{aligned}
\end{equation}
where we used:
\begin{equation}
 \begin{aligned}
  \langle P^{p,q},P^{p,q'}\rangle_{KG} &= -\langle \bar{P}^{p,q},\bar{P}^{p,q'}\rangle_{KG} &&= \delta_{qq'}N^2_{P^p}, \\
  \langle P^{p,q},\bar{P}^{p,q'}\rangle_{KG} &= \langle \bar{P}^{p,q},P^{p,q'}\rangle_{KG} &&= 0.
 \end{aligned}
\end{equation}
Using (\ref{alfabeta}) we find: 
\begin{equation}\label{aabb}
 \begin{aligned}
  \sum_{q=L,R}\left(\alpha^{\sigma}_{p,q}\bar{\alpha}^{\sigma}_{p,q}-\beta^{\sigma}_{p,q}\bar{\beta}^{\sigma}_{p,q}\right) &= \frac{8\sinh \pi p \left(\cosh \pi p -\sigma \cos \pi\nu'\right)}{\left|\Gamma[\nu'+ip+1]\right|^2} .
 \end{aligned}
\end{equation}
So finally we can substitute (\ref{aabb}) into (\ref{Nchi}):
\begin{equation}
 \begin{aligned}
  N^2_{\chi^{\sigma,p}} &= N^2_{P^p} \sum_{q=L,R}\left(\alpha^{\sigma}_{p,q}\bar{\alpha}^{\sigma}_{p,q}-\beta^{\sigma}_{p,q}\bar{\beta}^{\sigma}_{p,q}\right) \\
  &= \frac{2\sinh \pi p}{\pi}\times \frac{8\sinh \pi p \left(\cosh \pi p -\sigma \cos \pi\nu'\right)}{\left|\Gamma[\nu'+ip+1]\right|^2} \\
  &= \frac{16 \sinh^2 \pi p \left(\cosh \pi p -\sigma \cos \pi\nu'\right)}{\pi \left|\Gamma[\nu'+ip+1]\right|^2}.
 \end{aligned}
\end{equation}
This is consistent with \cite{Sasaki:1995}, but note that we included an extra factor of $2\sinh \pi p$ into the normalization, in order to simplify the expressions (\ref{alfabeta}). The normalized mode functions are the same as in \cite{Sasaki:1995}, of course.

\subsection{Mode functions for the massless scalar field}\label{app:modefunctionscalar}
Since we are mostly concerned with the massless minimally coupled scalar field ($\nu'=1$), we state the normalized mode functions for that case explicitly in hyperbolic time coordinate $t$ and in conformal time $\eta = \ln \tanh \frac{t}{2}$:
\begin{equation}\label{masslessmodes}
 \begin{aligned}
  \frac{1}{N_{P^p}}\frac{H}{\sinh t }P^{ip}_{1}(\cosh t) &= \frac{H}{\sqrt{2p(p^2+1)}}\left(\coth\frac{t}{2}\right)^{\frac{ip}{2}}\left(p \ \text{csch} \  t+ i\coth t\right)\\
  &= \frac{H}{\sqrt{2p(p^2+1)}}e^{-ip\eta}\left(p\sinh \eta-i\cosh \eta \right),
 \end{aligned}
\end{equation}
where we have chosen a convenient phase factor in the normalization, that does not affect the physics. The conformal time $\eta$ is defined as:
\begin{equation}
 \begin{aligned}
  ds^2 &= \frac{1}{H^2}\left(-dt^2+\sinh^2t\left(dr^2+\sinh^2r d\Omega^2_2\right)\right) \\
  &= \frac{\sinh^2(t(\eta))}{H^2}\left(-d\eta^2+dr^2+\sinh^2r d\Omega^2_2\right) \\
  \Rightarrow \eta&= \int \frac{dt}{\sinh t} \\
  &= \ln \tanh \frac{t}{2}.
 \end{aligned}
\end{equation}
Other useful relations between ``hyperbolic" time $t$ and ``conformal" time $\eta$ are:
\begin{equation}
 \begin{aligned}
  \sinh t &= -\frac{1}{\sinh \eta}, 
  &&\cosh t = -\coth \eta.
 \end{aligned}
\end{equation}
At early times $\eta \rightarrow -\infty$ the mode function for the massless minimally coupled scalar field (\ref{masslessmodes}) behaves like a positive energy mode function: 
\begin{equation}
 \begin{aligned}
   \frac{H}{\sqrt{2p(p^2+1)}}e^{-ip\eta}\left(p\sinh \eta-i\cosh \eta \right) &= \sinh\eta \left(\frac{H(p+i)}{\sqrt{2p(p^2+1)}}e^{-ip\eta}+O(e^{2\eta})\right) \\
   &\approx \sinh\eta \frac{H(p+i)}{\sqrt{2p(p^2+1)}}e^{-ip\eta}.
 \end{aligned}
\end{equation}
The infinite boost limit $\gamma\rightarrow \infty$ (\ref{fixedcoord}) corresponds to the large $t$ (or small $\eta$) and large momentum limit. In particular in terms of conformal time, the rescaling for small $\eta$ reads $\eta \rightarrow \eta \, e^{-\gamma}$, implying that the combination $p \, \eta$ is invariant in the limit. This gives 
\begin{equation}
 \begin{aligned}
   e^{-ip\eta}\left(p\sinh \eta-i\cosh \eta \right) &= p \sinh\eta \, e^{-ip\eta}\left(1-\frac{i}{p \eta}+O(\eta)\right)  \\
   &\approx e^{-ip\eta}\left(p\eta -i\right).
 \end{aligned}
\end{equation}
This is exactly the mode function for a massless scalar field in the flat de Sitter slicing (up to the appropriate normalization).

\section{Power spectra for the massless field}\label{app:powerspectramasslessfield}
\subsection{Direct calculation}
\textbf{Power spectrum in hyperbolic vacuum}\\
The power spectrum in the hyperbolic vacuum can be computed in a straightforward way
\begin{equation}
 \begin{aligned}
  \langle \Omega_{H}|\phi_p\phi_p'|\Omega_H\rangle &= \frac{H^2}{\sinh^2(t)}\sum_{lm}\sum_{l'm'}\frac{Y_{plm}Y^*_{p'l'm'}}{N_{P^p}N_{P^{p'}}}  \\
  &\times \langle \Omega_{H}|\left(\hat{b}_{plm}P^{ip}_{\nu'}+\hat{b}^{\dagger}_{plm}P^{-ip}_{\nu'}\right)\left(\hat{b}_{p'l'm'}P^{ip'}_{\nu'}+\hat{b}^{\dagger}_{p'l'm'}P^{-ip'}_{\nu'}\right)|\Omega_H\rangle \\
  &= \frac{H^2}{\sinh^2(t)}\sum_{lm}\sum_{l'm'}Y_{plm}Y^*_{p'l'm'}\frac{P^{ip}_{\nu'}P^{-ip'}_{\nu'}}{N_{P^p}N_{P^{p'}}} \langle \Omega_{H}|\hat{b}_{plm}\hat{b}^{\dagger}_{p'l'm'}|\Omega_H\rangle \ \text{using} \ \ \hat{b}_{plm}|\Omega_H\rangle = 0 \\
  &= \delta(p-p')\frac{H^2}{\sinh^2(t)}\sum_{lm} \left|Y_{plm}\right|^2\frac{\left|P^{ip}_{\nu'}\right|^2}{N^2_{P^p}} \ \text{using} \left[\hat{b}_{plm},\hat{b}^{\dagger}_{p'l'm'}\right] = \delta(p-p')\delta_{ll'}\delta_{mm'} \\
  &= \delta(p-p')\frac{H^2}{\sinh^2(t)}\frac{p^2}{2\pi^2}\frac{\left|P^{ip}_{\nu'}\right|^2}{N^2_{P^p}},
 \end{aligned}
\end{equation}
where we used the completeness relation for the $Y_{plm}$ in the last step. For the massless minimally coupled case $\nu' = 1$ we have:
\begin{equation}
 \begin{aligned}
 \langle \Omega_{H}|\phi_p\phi_p'|\Omega_H\rangle &=  \delta(p-p')\frac{H^2}{\sinh^2(t)}\frac{p}{4\pi^2}\frac{\cosh^2(t)+p^2}{(p^2+1)},
 \end{aligned}
\end{equation}
and for large $t\rightarrow \infty$
\begin{equation}
 \begin{aligned}
 \langle \Omega_{H}|\phi_p\phi_p'|\Omega_H\rangle &\rightarrow  \delta(p-p')\frac{H^2}{4\pi^2}\frac{p}{(p^2+1)}.
 \end{aligned}
\end{equation}
The power spectrum for the massless minimally coupled scalar field is given by:
\begin{equation}
 \begin{aligned}
\langle \Omega_{H}|\phi^2|\Omega_H\rangle &= \int dp \int dp' \langle \Omega_{H}|\phi_p\phi_p'|\Omega_H\rangle \\
&= \int d \ln p \frac{H^2}{4\pi^2}\frac{p^2}{p^2+1} \\
\Rightarrow \Delta^2_{\phi,H}(p) &= \frac{H^2}{4\pi^2}\frac{p^2}{p^2+1}.
 \end{aligned}
\end{equation}
\textbf{Power spectrum in Bunch--Davies vacuum}\\
The computation is similar to the previous case:
\begin{equation}\label{2ptdirect3}
 \begin{aligned}
 \langle \Omega_{BD}|\phi_p\phi_{p'}|\Omega_{BD}\rangle &= \frac{H^2}{\sinh^2 t}\sum_{lm l'm'}  \sum_{\sigma\sigma'} \frac{Y_{plm}Y^*_{p'l'm'}}{N_{\chi_{p,\sigma}}N_{\chi_{p',\sigma'}}} \\
 &\times \langle \Omega_{BD}|\left(\hat{a}_{\sigma plm}\chi_{p,\sigma}+ \hat{a}^{\dagger}_{\sigma plm}\bar{\chi}_{p,\sigma}\right)\left(\hat{a}_{\sigma'p'l'm'}\chi_{p',\sigma'}+\hat{a}^{\dagger}_{\sigma'p'l'm'}\bar{\chi
 }_{p',\sigma'}\right)|\Omega_{BD}\rangle \\
 &= \frac{H^2}{\sinh^2 t}\sum_{lm l'm'}   Y_{plm}Y^*_{p'l'm'} \sum_{\sigma\sigma'}\frac{\chi_{p,\sigma}\bar{\chi
 }_{p',\sigma'}}{N_{\chi_{p,\sigma}}N_{\chi_{p',\sigma'}}} \langle \Omega_{BD}|\hat{a}_{\sigma plm}\hat{a}^{\dagger}_{\sigma'p'l'm'}|\Omega_{BD}\rangle \\
 &= \delta(p-p')\frac{H^2}{\sinh^2 t}\sum_{lm}|Y_{plm}|^2\sum_{\sigma}\left|\frac{\chi_{p,\sigma}}{N_{\chi_{p,\sigma}}}\right|^2  \ \text{using} \ [\hat{a}_{\sigma plm},\hat{a}_{\sigma' p'l'm'}^{\dagger}] \\
 &= \delta_{\sigma \sigma'}\delta_{ll'} \delta_{mm'} \delta(p-p') \\
 &= \delta(p-p')\frac{H^2}{\sinh^2 t}\frac{p^2}{2\pi^2}\sum_{\sigma}\left(\begin{array}{l}
\left(\frac{\alpha_{p,L}^{\sigma}\bar{\alpha}^{\sigma}_{p,L}}{N^2_{\chi^{p\sigma}}}+\frac{\beta_{p,L}^{\sigma}\bar{\beta}^{\sigma}_{p,L}}{N^2_{\chi^{p,\sigma}}}\right)\left|P^{ip}_{\nu'}\right|^2 \\
+\frac{\alpha_{p,L}^{\sigma}\bar{\beta}_{p,L}^{\sigma}}{N^2_{\chi^{p\sigma}}}P^{ip}_{\nu'}P^{ip}_{\nu'}+\frac{\bar{\alpha}^{\sigma}_{p,L}\beta_{p,L}^{\sigma}}{N^2_{\chi^{p\sigma}}}P^{-ip}_{\nu'}P^{-ip}_{\nu'}                                            \end{array}\right).
\end{aligned}
\end{equation}
In the last step we used the completeness relation for $Y_{plm}$ and the expansion of $\chi$ in terms of the associated Legendre polynomials (\ref{tanakaexpans},\ref{alfabeta}). Here we will compute the spectrum for the massless scalar field $(\nu'=1)$. For the massless minimally coupled scalar ($\nu'=1$) the cross terms involving $P^{ip}P^{ip}$ and $P^{-ip}P^{-ip}$ vanish:
\begin{equation}
 \begin{aligned}
  \sum_{\sigma}\frac{\alpha_{p,L}^{\sigma}\bar{\beta}_{p,L}^{\sigma}}{
 N^2_{\chi^{p\sigma}}} &\propto \sum_{\sigma}\frac{(e^{\pi p}+\sigma)(e^{-\pi p}+\sigma)}{\cosh \pi p +\sigma} \propto \sum_{\sigma}\sigma =  0, 
\end{aligned}
\end{equation}
and similarly for the term involving $P^{-ip}P^{-ip}$. So for the massless minimally coupled case $(\nu'=1)$ we have:
\begin{equation}\label{2ptmasslessdirect}
 \begin{aligned}
 \langle \Omega_{BD}|\phi_p\phi_{p'}|\Omega_{BD}\rangle 
 &= \delta(p-p')\frac{H^2}{\sinh^2 t}\frac{p^2}{2\pi^2}\sum_{\sigma}
\left(\frac{\alpha_{p,L}^{\sigma}\bar{\alpha}^{\sigma}_{p,L}+\beta_{p,L}^{\sigma}\bar{\beta}^{\sigma}_{p,L}}{N^2_{\chi^{p,\sigma}}}\right)\left|P^{ip}_{1}\right|^2 \\
&= \delta(p-p')\frac{H^2}{\sinh^2 t}\frac{p^2}{2\pi^2}\sum_{\sigma}\frac{\pi}{16\sinh^2\pi p}\frac{(e^{\pi p}+\sigma)^2+(e^{-\pi p}+\sigma)^2}{\cosh\pi p+\sigma}\left|P^{ip}_{1}\right|^2 \\
 &= \delta(p-p')\frac{H^2}{\sinh^2 t}\frac{p^2}{2\pi^2}\frac{\pi \cosh(\pi p)}{2\sinh^2(\pi p)}\left|P^{ip}_1\right|^2 \\
 &= \delta(p-p')\frac{H^2}{\sinh^2(t)}\frac{p(\cosh^2(t)+p^2)}{4\pi^2(p^2+1)}\coth(\pi p).
 \end{aligned}
\end{equation}
For large $t\rightarrow \infty$ we have:
\begin{equation}
 \begin{aligned}
\langle \Omega_{BD}|\phi_p\phi_{p'}|\Omega_{BD}\rangle 
 &= \delta(p-p')\frac{H^2}{4\pi^2} \frac{p\coth(\pi p)}{p^2+1}  
 \end{aligned}
\end{equation}
and
\begin{equation}
 \begin{aligned}
  \langle \Omega_{BD}|\phi^2|\Omega_{BD}\rangle &= \int dp \int dp'  \langle \Omega_{BD}|\phi_p\phi_{p'}|\Omega_{BD}\rangle \\
  &= \frac{H^2}{4\pi^2}\int dp \frac{p \coth \pi p}{p^2+1} +\text{supercurvature modes} \\
  &= \frac{H^2}{4\pi^2}\int d\ln p  \ \ \frac{p^2 \coth \pi p}{p^2+1} +\text{supercurvature modes}.
 \end{aligned}
\end{equation}
The power spectrum is given by
\begin{equation}\label{BDdirect}
 \begin{aligned}
 \Delta^2_{\phi,BD}(p) &= \frac{H^2}{4\pi^2}\frac{p^2 \coth\pi p}{p^2+1}
 \end{aligned}
\end{equation}
which reduces for $p\gg 1$ to an approximately scale invariant spectrum:
\begin{equation}
 \begin{aligned}
 \Delta^2_\phi(p) &\approx \frac{H^2}{4\pi^2}.
 \end{aligned}
\end{equation}

\subsection{Reduced density matrix calculation}\label{app:reduceddensitymatrixspectrum}
In this section we derive the power spectrum in the Bunch-Davies state using an alternative method. We consider the reduced density matrix that remains after having traced out the degrees of freedom in the right hyperbolic patch. We find the same answer as in the direct calculation (\ref{BDdirect},\ref{2ptmasslessdirect}). 
The reduced density matrix for the left hyperbolic patch has been calculated by Maldacena and Pimentel \cite{MalPim:2012} and is given by:
\begin{equation}\label{dm}
 \begin{aligned}
 \hat{\rho}_{L,p,l,m} &= Tr_{H_R} \left\{ |\Omega_{BD} \rangle \langle \Omega_{BD} | \right\} \\
 &= (1-|\gamma_p)|^2)\sum_{n=0}^\infty |\gamma_p|^{2n} |n ; p,l,m\rangle
\langle n;  p,l,m |
 \end{aligned}
\end{equation}
where for the \emph{massless} scalar field $\gamma_p$ and $|n ; p,l,m\rangle$ are given by\footnote{For the massive scalar field Maldacena and Pimentel apply a Bogoliubov transformation on the set of $\hat{b}_{plm}$ operators to bring $\hat{\rho}_L$ in the form of (\ref{dm}).}:
\begin{equation}\label{gamma}
 \begin{aligned}
  \gamma_p(m=0) &= ie^{-\pi p} \\
 |n ; p,l,m\rangle &= \frac{(\hat{b}^{\dagger}_{plm})^n}{\sqrt{n!}} |\Omega_{H}\rangle.
 \end{aligned}
\end{equation}
For a point in the left hyperbolic wedge $x\in L$ the two point function is given by (\ref{Hexpansion}):

\begin{equation}
 \begin{aligned}
  \langle \Omega_{BD}|\phi_p\phi_{p'}|\Omega_{BD}\rangle &=
\text{Tr}_{\mathcal{H}_L}\left\{\phi_p\phi_{p'}\hat{\rho}_{L}\right\} \\
&= \delta(p-p')\frac{H^2}{\sinh^2(t)}\sum_{lm}|Y_{plm}|^{2} \frac{|P^{ip}_1|^2}{N^2_{P^p}} \\
&\times (1-|\gamma_p|^2)\sum_n |\gamma_p|^{2n}\left(2n+1\right) \\
&= \delta(p-p')\frac{H^2}{\sinh^2(t)}\frac{p^2}{2\pi^2} \frac{\cosh^2(t)+p^2}{2p(p^2+1)} \frac{1+|\gamma_p|^2}{1-|\gamma_p|^2} \\
&= \delta(p-p')\frac{H^2}{\sinh^2(t)}\frac{p}{4\pi^2} \frac{\cosh^2(t)+p^2}{(p^2+1)} \coth(\pi p)
 \end{aligned}
\end{equation}
which is equal to the result of the direct calculation (\ref{2ptmasslessdirect}). 

\section{Divergence of the energy momentum tensor}\label{EMT}
As is the case in the Fulling-Rindler vacuum, the energy momentum tensor diverges at the null boundary of the hyperbolic patch. One could construct lightcone coordinates $u = \eta-r$ and $v = \eta+r$ in order to calculate $T_{uu}$. Equivalently, we consider the leading divergence of $T_{tt}$ in the $t\rightarrow 0$ limit, which is more convenient. 
\begin{equation}
 \begin{aligned}
  T_{tt} &= (\partial_t \phi)^2-\frac{1}{2}g_{tt}g^{\sigma \rho}(\partial_{\sigma}\phi)(\partial_{\rho}\phi) 
 \end{aligned}
\end{equation}
For the massless case we have:
\begin{equation}
 \begin{aligned}
  \langle T_{tt} \rangle &= \frac{1}{2}\langle (\partial_t\phi)^2 + g^{rr}(\partial_r\phi)^2+g^{\theta\theta}(\partial_{\theta}\phi)^2+g^{\phi\phi}(\partial_{\phi\phi}\phi)^2\rangle. 
 \end{aligned}
\end{equation}
One can calculate this directly using the hyperbolic mode functions (\ref{Hmodes}) and the density matrix (\ref{dm}) for the Bunch-Davies expectation value $\langle T_{\mu\nu}\rangle_{BD}$. Equivalently, for the leading order term we can use the Wightman functions $G^{+}(x,x')$ as given in \cite{Sasaki:1995}:
\begin{equation}
 \begin{aligned}
  \langle (\partial_{t}\phi)^2 \rangle &= \lim_{t' \rightarrow t}\partial_{t}\partial_{t'}G^{+}(t,t'),
 \end{aligned}
\end{equation}
and similarly for the other coordinates. Note that the contribution of the supercurvature modes to the Wightman function only leads to subleading divergences\footnote{For $\nu' > 0$ the supercurvature mode contribution to the Wightman function is \cite{Sasaki:1995}: \begin{equation} G^{+}_{*}(t,t',\zeta) = \frac{H^2}{4\pi^{\frac{5}{2}}}\Gamma[-\nu'+1]\Gamma[\nu'+\frac{3}{2}]\frac{\sinh (\nu')\zeta}{\sinh \zeta}\left(\sinh t \sinh t'\right)^{\nu'-1}.\end{equation} For the minimally coupled massless case $\nu = \frac{3}{2}$ the supercurvature mode becomes time-independent. The contribution to the energy momentum tensor is of subleading order.}. The contribution of the subcurvature modes to the Wightman function for the massless $\nu' =1$ case is given by \cite{Sasaki:1995}:
\begin{equation}
 \begin{aligned}
  G^{+}(t,t',\zeta) &= \frac{H^2}{\sinh t \sinh t'}\frac{1}{8\pi^2}\int_{-\infty}^{\infty}dp \frac{\sin p \zeta}{\sinh \zeta}\frac{e^{\pi p}}{\sinh \pi p}\frac{(\cosh t+ip)(\cosh t'- ip)}{1+p^2}\left(\frac{\tanh \frac{t'}{2}}{\tanh\frac{t}{2}}\right)^{ip}, \\
  \zeta &= \cosh r \cosh r'-\sinh r\sinh r'\left(\cos \theta \cos\theta'+\sin \theta \sin\theta'\cos(\phi -\phi' )\right).
 \end{aligned}
\end{equation}
One can check the following:
\begin{equation}\label{factors}
 \begin{aligned}
  \langle (\partial_t\phi)^2 \rangle &= \lim_{t'\rightarrow t}\partial_{t}\partial_{t'}G^{+}(t,t',\zeta) &&= \frac{H^2}{4\pi^2}\int_0^{p_F}dp \ p(p^2+1)\coth(\pi p)\frac{1}{t^4}+O\left(\frac{1}{t^2}\right) \\
  \langle (\partial_r\phi)^2 \rangle &= \lim_{r'\rightarrow r}\partial_{r}\partial_{r'}G^{+}(t,t,\zeta) &&= \frac{H^2}{4\pi^2}\int_0^{p_F}dp \ p(p^2+1)\coth(\pi p)\frac{1}{t^2}+O\left(t^0\right) \\
  \langle (\partial_{\theta}\phi)^2 \rangle &= \lim_{\theta'\rightarrow \theta}\partial_{\theta}\partial_{\theta'}G^{+}(t,t,\zeta) &&=  \sinh^2r \frac{H^2}{4\pi^2}\int_0^{p_F}dp \ p(p^2+1)\coth(\pi p)\frac{1}{t^2}+O\left(t^0\right) \\
   \langle (\partial_{\phi}\phi)^2 \rangle &= \lim_{\phi'\rightarrow \phi}\partial_{\phi}\partial_{\phi'}G^{+}(t,t,\zeta) &&= \sin^2\theta \sinh^2r \frac{H^2}{4\pi^2}\int_0^{p_F}dp \ p(p^2+1)\coth(\pi p)\frac{1}{t^2}+O\left(t^0\right). 
 \end{aligned}
\end{equation}
Note that all these are divergent as $t\rightarrow 0$, but they also show the usual UV-divergence. The UV-divergence is regulated by a cutoff $ p_F$. The difference $\langle T_{tt}\rangle_H-\langle T_{tt}\rangle_{BD}$ will be UV-finite. We combine the components (\ref{factors}) to obtain $\langle T_{tt}\rangle_{BD}$. The expectation value $\langle T_{tt}\rangle_H$ is obtained by replacing $\coth \pi p \rightarrow 1$, where we use the expectation value $\langle \hat{b}^{\dagger}_{plm}\hat{b}_{plm}+\mathbf{I}\rangle$ in the two different states \footnote{We can calculate $\langle \hat{b}^{\dagger}_{plm}\hat{b}_{plm}+\mathbf{I}\rangle$ by using the density matrix (\ref{dm})}:
\begin{equation}
 \begin{aligned}
  \langle \hat{b}^{\dagger}_{plm}\hat{b}_{plm}+\mathbf{I}\rangle_{BD} &=\coth \pi p \\
  \langle \hat{b}^{\dagger}_{plm}\hat{b}_{plm}+\mathbf{I}\rangle_{H} &= 1.
 \end{aligned}
\end{equation}
Finally, we calculate the difference $\langle T_{tt}\rangle_H-\langle T_{tt}\rangle_{BD}$:
\begin{equation}
 \begin{aligned}
  \langle T_{tt}\rangle_H-\langle T_{tt}\rangle_{BD} &= \frac{H^4}{2\pi^2}\int_0^{\infty} dp \ p(p^2+1)\left(1-\coth \pi p\right) \frac{1}{t^4} + O\left(\frac{1}{t^2}\right) \\
  &= -\frac{11}{240\pi}\frac{1}{t^4}+O\left(\frac{1}{t^2}\right).
 \end{aligned}
\end{equation}
Note that we took the cutoff $p_F$ to infinity and obtain a UV-finite integral.

\section{The hyperbolic vacuum embedded in the Bunch--Davies state}\label{app:Hvacuum}
From \cite{MalPim:2012} we have for the massless case $\nu' = 1$:
\begin{equation}
 \begin{aligned}
  |\Omega_{\text{BD}}\rangle &= \left(\otimes_{plm} e^{\gamma_p \hat{b}_{Lplm}^{\dagger}\otimes \hat{b}^{\dagger}_{Rplm}}\right)|\Omega_{H,L}\rangle\otimes |\Omega_{H,R}\rangle
 \end{aligned}
\end{equation}
or suppressing the indices $p,l,m$:
\begin{equation}
 \begin{aligned}
  |\Omega_{\text{BD}}\rangle &= e^{\gamma \hat{b}_{L}^{\dagger}\otimes \hat{b}^{\dagger}_{R}}|\Omega_{H,L}\rangle\otimes |\Omega_{H,R}\rangle
 \end{aligned}
\end{equation}
with $\gamma = ie^{-\pi p}$.
The left hyperbolic vacuum $|\Omega_{H,L}\rangle$ is not a state of the full system; we need information about the state in the right hyperbolic patch as well. The simplest way to embed the left hyperbolic vacuum in the full Hilbert space, we can consider the simple and symmetric state $|\Omega_{H,L}\rangle\otimes |\Omega_{H,R}\rangle$. This state is \emph{not} the natural vacuum state (the Bunch--Davies state) for the full de Sitter space. \\
\textbf{Proposition:}
\begin{equation}\label{prop}
 \begin{aligned}
  |\Omega_{H,L}\rangle\otimes |\Omega_{H,R}\rangle &\propto  e^{-|\gamma_p| \ \hat{a}^{\dagger}_{+}\otimes  \ \hat{a}^{\dagger}_{-}}|\Omega_{BD}\rangle.
 \end{aligned}
\end{equation}
\textbf{Proof:} \\
We will show that the right hand side of (\ref{prop}) vanishes when we act with any of the $\hat{b}_{Llpm}$ annihilation operators. We use the expression for the hyperbolic annihilation operator $\hat{b}_{Llpm}$ in terms of the creation and annihilation for Bunch-Davies modes (\ref{relationab}), suppressing from now on the labels $p,l,m$:
\begin{equation}\label{basimple}
 \begin{aligned}    
\hat{b}_{L} &= \sum_{\sigma}\frac{N_{P}}{N_{\chi_{\sigma}}}\left(\alpha^{\sigma}_{L}\hat{a}_{\sigma}+\bar{\beta}^{\sigma}_{L}\hat{a}^{\dagger}_{\sigma}\right).
 \end{aligned}
\end{equation}
We want to show that $\hat{b}_{L}$ acting on the RHS of (\ref{prop}) vanishes:
\begin{equation}\label{duurtlang}
 \begin{aligned}
  \sum_{\sigma}\frac{N_{P}}{N_{\chi_{\sigma}}}\left(\alpha^{\sigma}_{L}\hat{a}_{\sigma}+\bar{\beta}^{\sigma}_{L}\hat{a}^{\dagger}_{\sigma}\right) e^{-|\gamma_p| \hat{a}^{\dagger}_{+}\otimes \hat{a}^{\dagger}_{-}}|\Omega_{BD}\rangle \stackrel{?}{=} 0.
 \end{aligned}
\end{equation}
Consider the annihilation operator $\hat{a}_{\sigma}$ acting on (\ref{prop}):
\begin{equation}
 \begin{aligned}
  \hat{a}_{\pm}e^{-|\gamma_p|\hat{a}^{\dagger}_{+}\hat{a}^{\dagger}_{-}}|\Omega_{BD}\rangle &= \left[\hat{a}_{\pm} \ , \ e^{-|\gamma_p| \hat{a}^{\dagger}_{+}\hat{a}^{\dagger}_{-}}\right]|\Omega_{BD}\rangle \\
  &= -\left[\hat{a}_{\pm} \ , \ |\gamma_p|\hat{a}^{\dagger}_{+}\hat{a}^{\dagger}_{-} \right] e^{-|\gamma_p| \hat{a}^{\dagger}_{+}\hat{a}^{\dagger}_{-}}|\Omega_{BD}\rangle \\
  &= -|\gamma_p| \hat{a}^{\dagger}_{\mp} e^{-|\gamma_p| \hat{a}^{\dagger}_{+}\hat{a}^{\dagger}_{-}}|\Omega_{BD}\rangle. 
 \end{aligned}
\end{equation}
Substituting this result in (\ref{duurtlang}) gives:
\begin{equation}\label{duurtlanger}
 \begin{aligned}
  \sum_{\sigma}\frac{N_{P}}{N_{\chi_{\sigma}}}\left(\alpha^{\sigma}_{L}\hat{a}_{\sigma}+\bar{\beta}^{\sigma}_{L}\hat{a}^{\dagger}_{\sigma}\right) e^{|\gamma_p| \hat{a}^{\dagger}_{+}\otimes \hat{a}^{\dagger}_{-}}|\Omega_{BD}\rangle =  N_{P}\sum_{\sigma}\left(-\frac{\alpha^{-\sigma}_{L}}{N_{\chi_{-\sigma}}}|\gamma|+\frac{\bar{\beta}^{\sigma}_{L}}{N_{\chi_{\sigma}}}\right)\hat{a}^{\dagger}_{\sigma} e^{|\gamma_p| \hat{a}^{\dagger}_{+}\otimes \hat{a}^{\dagger}_{-}}|\Omega_{BD}\rangle.
 \end{aligned}
\end{equation}
It is easy to check that the quantity between brackets on the RHS of (\ref{duurtlanger}) vanishes for both $\sigma = \pm 1$. This finalizes the proof:
\begin{equation}
 \hat{b}_{L}e^{-|\gamma|\hat{a}^{\dagger}_{+}\otimes \hat{a}^{\dagger}_{-}}|\Omega_{BD}\rangle = 0 \ \ \ \ \ \ \  \forall p,l,m.
\end{equation}

The state (\ref{prop}) is pure. Note that the symmetric and antisymmetric modes corresponding to $\sigma=\pm1$ are entangled with each other and their reduced density matrices are thermal.

\end{appendix}


\begin{thebibliography}{999}
   

 
\bibitem{Planck:2013}
  P.~A.~R.~Ade {\it et al.}  [Planck Collaboration],
  {\em Planck 2013 results. XXII. Constraints on inflation},
  Astron.\ Astrophys.\  {\bf 571} (2014) A22
  [arXiv:1303.5082 [astro-ph.CO]].
  
\bibitem{BauMcA:2014}
  D.~Baumann and L.~McAllister,
  {\em Inflation and String Theory},
  arXiv:1404.2601 [hep-th].
  
\bibitem{Vilenkin:1983}
  A.~Vilenkin,
  {\em The Birth of Inflationary Universes},
  Phys.\ Rev.\ D {\bf 27} (1983) 2848.
  
\bibitem{Susskind:2003}
  L.~Susskind,
  {\em The Anthropic landscape of string theory},
  In *Carr, Bernard (ed.): Universe or multiverse?* 247-266
  [hep-th/0302219].

\bibitem{Guth:2012}
  A.~H.~Guth and Y.~Nomura,
  {\em What can the observation of nonzero curvature tell us?},
  Phys.\ Rev.\ D {\bf 86} (2012) 023534
  [arXiv:1203.6876 [hep-th]].

\bibitem{Freivogel:2006}
B.~Freivogel, M.~Kleban, M.~Rodriguez Martinez and L.~Susskind,
  {\em Observational consequences of a landscape},
  JHEP {\bf 0603} (2006) 039
  [hep-th/0505232].
  
\bibitem{YamLinNarSasTan:2011}
  D.~Yamauchi, A.~Linde, A.~Naruko, M.~Sasaki and T.~Tanaka,
  {\em Open inflation in the landscape},
  Phys.\ Rev.\ D {\bf 84} (2011) 043513
  [arXiv:1105.2674 [hep-th]].

\bibitem{BouHarSen:2013}
  R.~Bousso, D.~Harlow and L.~Senatore,
  {\em Inflation after False Vacuum Decay: Observational Prospects after Planck},
  arXiv:1309.4060 [hep-th].

\bibitem{Kleban:2009}
B.~Freivogel, M.~Kleban, A.~Nicolis and K.~Sigurdson,
  {\em Eternal Inflation, Bubble Collisions, and the Disintegration of the Persistence of Memory},
  JCAP {\bf 0908} (2009) 036
  [arXiv:0901.0007 [hep-th]]. 

\bibitem{Kleban:2011}
  M.~Kleban,
  {\em Cosmic Bubble Collisions},
  Class.\ Quant.\ Grav.\  {\bf 28} (2011) 204008
  [arXiv:1107.2593 [astro-ph.CO]].
  
\bibitem{Kleban:2012}
  R.~Gobbetti and M.~Kleban,
  {\em Analyzing Cosmic Bubble Collisions},
  JCAP {\bf 1205} (2012) 025
  [arXiv:1201.6380 [hep-th]].

\bibitem{Peiris:2011}
S.~M.~Feeney, M.~C.~Johnson, D.~J.~Mortlock and H.~V.~Peiris,
  {\em First Observational Tests of Eternal Inflation},
  Phys.\ Rev.\ Lett.\  {\bf 107} (2011) 071301
  [arXiv:1012.1995 [astro-ph.CO]]. 
 
\bibitem{Kanno:2014}
  S.~Kanno,
  {\em Impact of quantum entanglement on spectrum of cosmological fluctuations},
  JCAP {\bf 1407} (2014) 029
  [arXiv:1405.7793 [hep-th]]. 

\bibitem{Komatsu:2013}
  K.~Sugimura and E.~Komatsu,
  {\em Bispectrum from open inflation},
  JCAP {\bf 1311} (2013) 065
  [arXiv:1309.1579 [astro-ph.CO]].

\bibitem{Agullo:2010}
  I.~Agullo and L.~Parker,
  {\em Non-gaussianities and the Stimulated creation of quanta in the inflationary universe},
  Phys.\ Rev.\ D {\bf 83} (2011) 063526
  [arXiv:1010.5766 [astro-ph.CO]].

\bibitem{ColLuc:1980}
  S.~R.~Coleman and F.~De Luccia,
  {\em Gravitational Effects on and of Vacuum Decay},
  Phys.\ Rev.\ D {\bf 21} (1980) 3305.

\bibitem{Gott:1982}
  J.~R.~Gott,
  {\em Creation of Open Universes from de Sitter Space},
  Nature {\bf 295} (1982) 304.


  
\bibitem{BucGolTur:1994}
  M.~Bucher, A.~S.~Goldhaber and N.~Turok,
  {\em An open universe from inflation},
  Phys.\ Rev.\ D {\bf 52} (1995) 3314
  [hep-ph/9411206].
  
\bibitem{LytSte:1990}
D.~H.~Lyth, E.~D.~Stewart,
{\em Inflationary density perturbations with $\Omega <1$},
Phys. \ Lett. B {\bf 252} (1990) 336
 
 
\bibitem{Sasaki:1995} M. Sasaki, T. Tanaka and K. Yamamoto, {\em Euclidean vacuum mode functions for a scalar field on open de Sitter space}, Phys. Rev. D {\bf 51}, 2979 (1995), arxiv:gr-qc/9412025 

\bibitem{MalPim:2012} J. Maldacena and G.L. Pimentel, {\em Entanglement entropy in de Sitter space}, JHEP {\bf 1302}, 038 (2013), arXiv:1210.7244[hep-th].

\bibitem{GreParScha:2006}
  B.~Greene, M.~Parikh and J.~P.~van der Schaar,
  {\em Universal correction to the inflationary vacuum},
  JHEP {\bf 0604} (2006) 057
  [hep-th/0512243].

\bibitem{ParVer:2004}
  M.~K.~Parikh and E.~P.~Verlinde,
  {\em De Sitter holography with a finite number of states},
  JHEP {\bf 0501} (2005) 054
  [hep-th/0410227].

\bibitem{Cohn:1998}
  J.~D.~Cohn and D.~I.~Kaiser,
  {\em Where do the supercurvature modes go?},
  Phys.\ Rev.\ D {\bf 58} (1998) 083515
  [gr-qc/9803073].

\bibitem{Parentani:1993}
  R.~Parentani,
  {\em The Energy momentum tensor in Fulling-Rindler vacuum},
  Class.\ Quant.\ Grav.\  {\bf 10} (1993) 1409
  [hep-th/9303062].

\bibitem{Eme:2014}
  S.~Emelyanov,
  {\em Local thermal observables in spatially open FRW spaces},
  arXiv:1406.3360 [gr-qc].  
  
\bibitem{Sasaki:1996} K.~Yamamoto, M.~Sasaki and T.~Tanaka, {\em Quantum fluctuations and CMB anisotropies in one bubble open inflation models,} {\em Phys.~Rev.} {\bf D54}(1996) 5031--5048, [astro-ph/9605103]

\bibitem{Gar:1998}
  J.~Garriga, X.~Montes, M.~Sasaki and T.~Tanaka,
  {\em Spectrum of cosmological perturbations in the one bubble open universe},
  Nucl.\ Phys.\ B {\bf 551} (1999) 317
  [astro-ph/9811257].

\bibitem{HawHerTur:2000}
  S.~W.~Hawking, T.~Hertog and N.~Turok,
  {\em Gravitational waves in open de Sitter space},
  Phys.\ Rev.\ D {\bf 62} (2000) 063502
  [hep-th/0003016].
  
  \bibitem{Holman:2007}
  R.~Holman and A.~J.~Tolley,
  {\em Enhanced Non-Gaussianity from Excited Initial States},
  JCAP {\bf 0805} (2008) 001
  [arXiv:0710.1302 [hep-th]].
  
  \bibitem{Meerburg:2009}
  P.~D.~Meerburg, J.~P.~van der Schaar and P.~S.~Corasaniti,
  {\em Signatures of Initial State Modifications on Bispectrum Statistics},
  JCAP {\bf 0905} (2009) 018
  [arXiv:0901.4044 [hep-th]].

\bibitem{Holman:2012} N.~Agarwal, R.~Holman, A.J.~Tolley and J.ZLin, {\em Effective field theory and non gaussianity from general initial states}, arXiv:1212.1172[hep-th]

\bibitem{Maldacena:2002} J.~Maldacena, {\em Non Gaussian features of primordial fluctuations in single field inflationary models }, [astro-ph/0210603v5]




 
\bibitem{Magnus} The Legendre function $P$ and $Q$ correspond to
$\cal B$ and $\cal D$, respectively, in a book by
  W. Magnus, F. Oberhettinger and R. P. Soni, {\it Formulas and
 Theorems for the Special Functions of Mathematical  Physics},
 (Springer-Verlag, Berlin Heidelberg, 1966).  



  
\end{thebibliography}
\end{document}